\documentclass[a4paper,twocolumn,11pt,accepted=2017-05-09]{quantumarticle}
\pdfoutput=1
\usepackage[utf8]{inputenc}
\usepackage[english]{babel}
\usepackage[T1]{fontenc}
\usepackage{amsmath}
\usepackage{hyperref}
\usepackage{amssymb}
\usepackage{float}
\usepackage{makecell}
\usepackage{algorithm} 
\usepackage{algorithmicx}
\usepackage{algpseudocode}
\usepackage{booktabs}

\usepackage{tikz}
\usepackage{lipsum}

\begin{document}

\title{A Quantum Neural Network-Based Approach to Power Quality Disturbances Detection and Recognition}

\author{Guo-Dong Li} 
\affiliation{School of Control and Computer Engineering, North China Electric Power University, Beijing 102206, China}
\author{Hai-Yan He}
\affiliation{School of Control and Computer Engineering, North China Electric Power University, Beijing 102206, China}
\author{Yue Li}
\affiliation{lectric Power Science Research Institute, Guizhou Power Grid Co., Ltd., Guiyang 550005, Guizhou
Province, China}
\author{Xin-Hao Li}
\affiliation{lectric Power Science Research Institute, Guizhou Power Grid Co., Ltd., Guiyang 550005, Guizhou
Province, China}
\author{Hao Liu}
\affiliation{School of Electrical and Electronic Engineering, North China Electric Power University, Beijing 102206, China}
\author{Qing-Le Wang*}
\email{wqle519@gmail.com}
\affiliation{School of Control and Computer Engineering, North China Electric Power University, Beijing 102206, China}
\author{Long Cheng}
\affiliation{School of Control and Computer Engineering, North China Electric Power University, Beijing 102206, China}
\maketitle

\begin{abstract}
Power quality disturbances (PQDs) significantly impact the stability and reliability of power systems, necessitating accurate and efficient detection and recognition methods. While numerous classical algorithms for PQDs detection and recognition have been extensively studied and applied, related work in the quantum domain is still in its infancy. In this paper, an improved quantum neural networks (QNN) model for PQDs detection and recognition is proposed. Specifically, the model constructs a quantum circuit comprising data qubits and ancilla qubits. Classical data is transformed into quantum data by embedding it into data qubits via the encoding layer. Subsequently, parametric quantum gates are utilized to form the variational layer, which facilitates qubit information transformation, thereby extracting essential feature information for detection and recognition. The expected value is obtained by measuring ancilla qubits, enabling the completion of disturbance classification based on this expected value. An analysis reveals that the runtime and space complexities of the QNN are $O\left ( poly\left ( N \right )  \right )$ and  $O\left ( N \right )$, respectively. Extensive experiments validate the feasibility and superiority of the proposed model in PQD detection and recognition. The model achieves accuracies of 99.75\%, 97.85\% and 95.5\% in experiments involving the detection of disturbances, recognition of seven single disturbances, and recognition of ten mixed disturbances, respectively. Additionally, noise simulation and comparative experiments demonstrate that the proposed model exhibits robust anti-noise capabilities, requires few training parameters, and maintains high accuracy.

\end{abstract}

\section{Introduction}
Power quality (PQ) serves as a pivotal indicator for evaluating the operational efficiency of the power system, which is directly related to the stable operation of the power system and the user's experience of electricity consumption. The rapid evolution of smart grids has facilitated the widespread integration of various types of nonlinear loads into power systems, resulting in increasingly intricate grid signals and consequently triggering a series of PQ issues\cite{ref1}. The utilization of new energy vehicle charging stations, power transfer switches, distributed energy sources, and various reactive devices often entails instability and intermittency, leading to the generation of disturbance signals within the power system. These signals manifest as power quality disturbances (PQDs), characterized by fluctuations in voltage, current, and frequency\cite{ref2}. Furthermore, the integration of renewable energy (RE) sources into the power system can exacerbate PQ issues\cite{ref3}. The inherent instability of RE generation serves as a significant catalyst for PQDs, leading to voltage imbalances or fluctuations\cite{ref4}. These issues pose a threat to the security and stability of the power system, while also resulting in significant economic losses for grid companies and users. In order to improve PQ, it is necessary to detect, recognize and classify PQDs before taking solutions. Therefore, PQDs detection and recognition is the premise of solving PQ issues.

With the increasing importance of PQDs, researchers have conducted extensive research on the detection and recognition of PQDs. Generally, the detection and recognition process of PQDs can be divided into three stages: signal analysis, feature extraction and intelligent classification\cite{ref5}. In recent years, signal processing techniques have emerged as a common approach for analyzing PQDs signals to extract crucial disturbance features. Methods such as short-time Fourier transform (STFT)\cite{ref6}, wavelet transform (WT)\cite{ref7}, and S-transform (ST)\cite{ref8} have been instrumental in capturing characteristic changes in non-stationary signals by tracking local time-frequency variations. Among them, ST is the inheritance and development of STFT and WT, while avoiding some shortcomings of both. Notably, ST exhibits the ability to detect interference amidst noise, making it widely adopted in PQDs analysis\cite{ref9}. Regarding classification and recognition, advancements in computing power have spurred the continual proposal of PQDs classification methods based on pattern recognition such as decision tree (DT)\cite{ref10}, support vector machine (SVM)\cite{ref11}, neural network-based methods and so on. In 2020, Khoa et al. combine ST and rule-based decision DT to recognize PQDs\cite{ref12}. In 2020, Tang et al. used hyperbolic ST and SVM to classify PQDs events in distributed generation systems\cite{ref13}. In 2023, a hybrid method called PQDs assessment of PQEventCog was proposed by Fu et al.\cite{ref14}, which uses optimized ST and convolutional neural network (CNN) to classify PQDs. In 2023, Abubakar et al. proposed a new algorithm based on discrete orthogonal ST and compressed neural network to realize high accuracy recognition of PQDs \cite{ref21}. The continuous development of PQDs detection and recognition methodologies underscores the importance of addressing PQ concerns in contemporary power systems.

The rise of quantum computing heralds quantum machine learning (QML) as an innovative and promising field\cite{ref15}. Among the many areas of QML, quantum neural networks (QNN) combine the respective strengths of quantum computing and neural network models to enable more powerful artificial intelligence. The advantages of QNN are mainly reflected in the ability of parallel processing of quantum computing to obtain quantum acceleration\cite{ref16}, extract the optimal network architecture\cite{ref17} at a lower cost, and improve the accuracy, generalization and robustness of the network\cite{ref18}. As early as 1995, the concept of QNN was first proposed\cite{ref19}. After that, the researchers have begun further studies on QNN. In 2021, Abbas et al. demonstrated for the first time that well-designed QNN have higher effective dimensions and faster training capabilities than classical neural networks\cite{ref20}. In 2022, Stein et al. proposed a novel architecture, QuClassi, which is a QNN for both second-class and multi-class classification\cite{ref47}. Compared to classical deep neural networks with similar performance, Quclassi is able to reduce the parameters by 97.37\%. QNN has been successful in a variety of tasks, including classification\cite{ref47}, regression\cite{ref22} and reinforcement learning\cite{ref48}. Over the years, QNN has made extraordinary advances that have had a huge impact on computing needs, offering possibilities for solving problems that classical computers cannot solve. 

By harnessing the complex properties of quantum states and the adaptability of quantum circuits, QML holds promise to enhance the capability to address PQDs. Recognizing the potential of QML, we propose an improved QNN model for PQDs detection and recognition. The model adopts a hybrid classical-quantum method, and uses the limited qubits to construct the quantum circuit that meets requirements, which effectively reduces the complexity and resources of implementation. Overall, the main contributions of this paper are as follows:

(1) In this paper, we propose an improved QNN model for PQDs detection and recognition in the power field. On the original basis, the QNN model can be extended to binary and multi-classification models for disturbance detection and recognition. To the best of our knowledge, this is the first attempt to apply QNN to the field of PQDs.

(2) The model uses data and ancilla qubits to build strongly entangled quantum circuit. At the same time, the model adopts hierarchical design to provide detailed description for each layer. The model maintains high classification and recognition accuracy with fewer training parameters and shallower circuit depth, significantly reducing the consumption of computational resources.

(3) Extensive experiments are conducted to verify the feasibility of the proposed method for PQDs detection and recognition. An accuracy of 99.75\% is achieved in the experiment for detecting the presence or absence of disturbances, while accuracies of 97.85\% and 95.5\% are respectively achieved in the subsequent classification experiments for identifying single and mixed PQDs. In addition, our proposed QNN model shows its outstanding advantages in comparison with the rest of classical methods.

(4) The proposed method has good anti-noise ability in noisy environment. By performing seven classification experiments for recognising single PQDs under noise conditions with signal-to-noise ratios (SNR) of 20dB, 30dB and 40dB, the results show that the recognition accuracies are all above 96\%.

The rest of the paper is organized as follows: The relevant knowledge is introduced in section 2. In section 3, the QNN-based PQDs detection and recognition model is presented, and its detailed design is described from several aspects. Various experiments are analyzed and discussed in section 4. Finally, we summarize the work of this paper in section 5.

\section{Related Knowledge} 
\subsection{Preliminaries of Quantum Computing}
In quantum computing, all computations are reduced to a series of quantum bits (qubits) operations involving rotation and entanglement. In a classical computer, a classical bit can only represent 0 or 1. However, a qubit can be state $\left|0\right\rangle$, state $\left|1  \right\rangle$, or a superposition of $\left|0\right\rangle$ and $\left|1\right\rangle$. Any quantum state of a qubit $\left|\psi\right\rangle$ can represent a linear combination of $\left|0\right\rangle$ and $\left|1\right\rangle$, i.e:
\begin{equation}
\left|\psi\right\rangle=\alpha\left|0\right\rangle+\beta\left|1\right\rangle=
\begin{bmatrix}
 \alpha \\ \beta 
  \end{bmatrix}
\end{equation}
where $\alpha$ and $\beta$ are complex numbers known as probability amplitudes that satisfy the normalization condition $\left|\alpha\right|^{2}+\left|\beta\right|^{2}=1$. Fig.1 shows a single qubit using a Bloch sphere as the basic unit of quantum information. It represents the quantum state of a qubit as a dot on its surface, each dot describing a unique superposition and phase state.
\begin{figure}[h]
		\centering
		\includegraphics[width=0.7\linewidth]{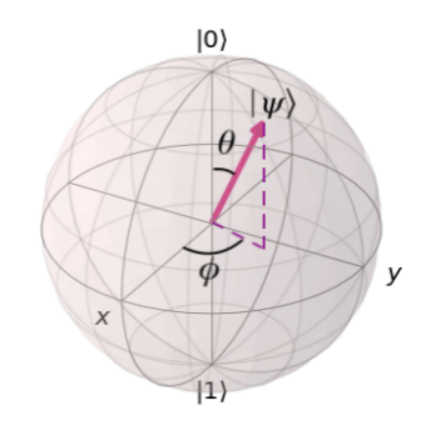}
		\caption{Bloch spherical representation of a qubit\cite{ref49} }
		\label{FIG:1}
\end{figure}

In quantum computing, transformations on qubits are implemented by controlled rotations represented by unitary operators:
\begin{equation}
    \left|\psi\right\rangle = R_{y}\left(\theta_{y}\right)\left|0\right\rangle
\end{equation} 
Where $R_{y}\left(\theta_{y}\right)$ is the Pauli operator that causes qubit rotation on the y-axis. The resulting qubit rotations are parameterized with $\theta _{y}$, similar to the weights in a classical neural network. In the context of the Bloch sphere, this rotational operation can put a qubit in a superposition state between $\left|0\right\rangle$ and $\left|1\right\rangle$. We denote $R_{y}$ gate as:
\begin{equation}
    R_{y}\left(\theta\right)=\begin{bmatrix}
 \cos\frac{\theta }{2}  & - \sin\frac{\theta }{2}  \\
 \sin\frac{\theta }{2}  & \cos\frac{\theta }{2} 
\end{bmatrix}
\end{equation}

In contrast, entanglement operations can occur between multiple qubits. Considering two arbitrary qubits $\left|\phi_{1}\right\rangle$ and $\left|\phi_{2}\right\rangle$, the Controlled-$R_{y}(CR_{y})$ gate applies its unique unitary operation to compute their joint state, which can be expressed as:
\begin{equation}
\left|\phi_{1}\phi_{2}\right\rangle^{'}=CR_{y}\left(\theta\right )\left| \phi_{1}\phi_{2}\right \rangle
\end{equation}
The operation of the $CR_{y}$ gate can also be described in terms of a matrix, denoted as:
\begin{equation}
\begin{split}
    CR_{y}\left(\theta\right) &= I\otimes\left|0\right\rangle\left\langle0\right|+R_{y}\left(\theta  \right)\otimes\left|1\right\rangle\left\langle1\right|\\ &= \begin{bmatrix}
 1 & 0 & 0 & 0\\
 0 & \cos\frac{\theta}{2}  & 0 & -\sin\frac{\theta}{2} \\
 0 & 0 & 1 & 0\\
 0 & \sin\frac{\theta}{2}  & 0 & \cos\frac{\theta}{2} 
\end{bmatrix}
\end{split}
\end{equation}
\subsection{Quantum Neural Network}
QNN are one of the best-known algorithms in QML, integrating the concepts of quantum computing and artificial neural networks(ANN) for a variety of classification problems. Existing research methods on QNN include two main approaches: one imitates the linear and nonlinear operations of ANN; the other uses PQC as trainable neurons and mimics the hierarchical structure of ANN\cite{ref23,ref24}. In recent years, scholars have extensively explored a variety of QNN models\cite{ref25,ref26} based on medium-sized quantum processors, such as quantum perceptron\cite{ref27}, quantum tensor neural network(QTNN)\cite{ref28}, quantum convolutional neural network(QCNN)\cite{ref29} and so on. These quantum neural network models simulate classical quantum systems with network structure characteristics in quantum Hilbert space, and build different quantum circuits to approximate nonlinear functions. Landman et al. introduced two new quantum methods of neural networks, namely quantum assisted neural network and quantum orthogonal neural network\cite{ref30}. The effect of different architectures and training methods on classification accuracy is investigated through medical image classification experiments. Huang et al. proposed a classifier based on variational quantum tensor network (VQTN)\cite{ref31}. The method improves the classification accuracy by designing shallow quantum circuits and multiple reading methods. Li et al. proposed a more complex quantum convolutional neural network model\cite{ref32}, which uses a more refined quantum circuit to achieve quantum inner product computation and approximate nonlinear mapping of activation functions. A quantum federated learning (QFL) framework specifically designed for classical clients, dubbed CCQFL, has been proposed by Song et al.\cite{ref33} and its feasibility has been verified by a handwritten digit classification task. Since QNN are hybrid quantum-classical models, the training and optimization of QNN combines quantum and classical methods. 
\begin{figure*}[t]
	\centering
		\includegraphics[width=1.0\linewidth]{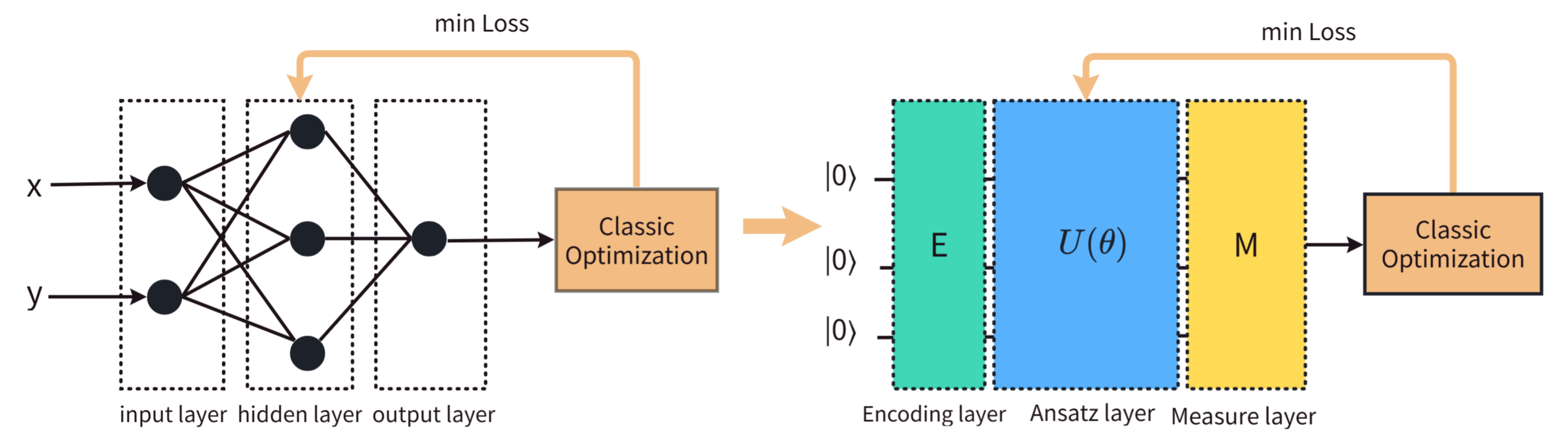}
	\caption{Classical neural network and quantum neural network structure diagram.According to classical neural networks, each layer of quantum neural networks can be mapped to classical neural networks.}
	\label{FIG:2}
\end{figure*} 

The common QNN structure is shown in Fig.2. Compared with artificial neural networks, the data encoding layer of QNN is equivalent to the input layer, encoding classical data into quantum states, and the specific encoding method is selected according to the actual situation. The quantum transformation layer is equivalent to the hidden layer. The quantum transformation layer adopts PQC designed with parametric quantum gates to transform and process the input quantum state, which involves the weight of parametric quantum gate equivalent to the traditional neural network. The quantum measurement layer is equivalent to the output layer, which can get the measurement value of the final quantum state. Among them, the classical optimizer is used to optimize and update the parameters of the quantum gate to get a model that meets the requirements. Research shows that compared with traditional neural networks, QNN based on PQC is more expressive in terms of parameters.
\section{PQDs Detection and Recognition Based on QNN}
In this section, we present the QNN model for detecting and recognizing PQDs. Firstly, the system framework and overall problem statement for solving PQDs are presented, in which the detection and recognition process of PQDs are elaborated. Next, the generation of PQDs and related data processing are presented. Subsequently, a comprehensive overview of the overall structure and hierarchical organisation of the proposed QNN model are presented.
\subsection{Detection and Recognition System Framework}
In this section, we present the detailed system framework, problem statement under PQDs detection and recognition.
\subsubsection{System Framework}
Fig.3 shows the system framework workflow for detecting and recognizing PQDs. The original disturbance waveform data are obtained by mathematical model, which contain a large number of data elements. These data may increase the complexity and calculation cost of the subsequent model. Therefore, we use feature extraction method to extract features with strong correlation. Each PQDs has unique characteristics and there are differences in the categorization criteria on which they are based. Nevertheless, each PQDs data is divided into the correct categories as much as possible. Due to the similarity between few features, there are errors in recognizing PQDs. Therefore, we update the model parameters based on the error so that its performance is improved. In this work, we aim to minimize the error between the true and predicted labels of PQDs. It will change the parameters of the model and improve the accuracy of classification. The final classification result can be obtained by performing PQDs classification on the model that determines the optimal parameters.
\begin{figure*}[t]
	\centering
		\includegraphics[width=1.0\linewidth]{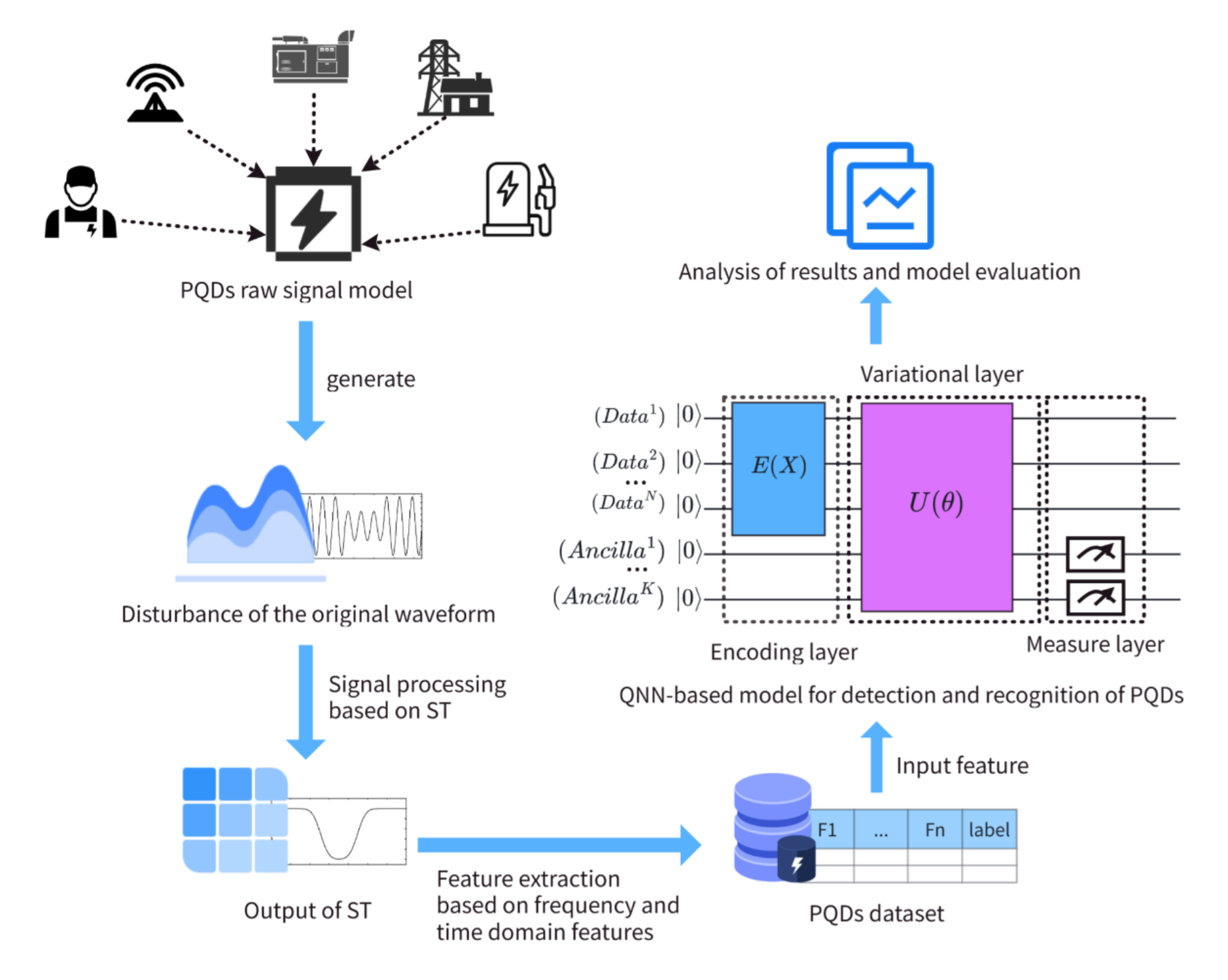}
	\caption{PQDs Detection and Recognition System Framework Workflow.}
	\label{FIG:3}
\end{figure*} 
\subsubsection{Problem Statement}
For the disturbance signals present in the power system, methods are taken to detect the disturbance signals therein and the abnormal signals are recognized to obtain the type of disturbance. The PQDs detection and recognition problem can be modelled as a classification problem, where PQDs detection is the classification of data into normal and abnormal signals and PQDs recognition is the detailed classification of abnormal signals into specific disturbance types. The problem we solve is described as follows:
we use equipment to obtain raw PQDs waveform data, which we express as:
\begin{equation}
    D=\left\{ w,y\right\} 
\end{equation}
where $y$ is the disturbance type label, $w$ is the raw PQDs waveform data.

The feature vector $X$ is extracted from the original waveform data $w$ using the feature extraction and selection function $E\left ( x \right )$ before classification:
\begin{equation}
    X=E\left ( w\right )
\end{equation}
where $X=\left [ x_{1},\dots ,x_{n} \right ]$, $x_{i}$ denotes the ith feature in the PQDs feature vector.

When QNN is applied to PQDs problem, the eigenvector $X$ of PQDs is first converted to quantum state into the subsequent quantum circuit through the encoding layer (E). The feature vector $X$ is embedded to obtain $\left | x  \right \rangle$:
\begin{equation}
    X\to \left | x  \right \rangle = E\left ( X  \right ) \left | 0  \right \rangle  
\end{equation}

Following the analogy with the classical case, a variational layer (U) can be defined as a parameterised circuit acting on the input state $\left | x  \right \rangle$ and producing the output state $\left | y  \right \rangle$:
\begin{equation}
    \left | x \right \rangle\to \left | y  \right \rangle = U\left ( \theta \right )\left | x  \right \rangle   
\end{equation}
where $\theta$ is an array of classical variational parameters.

Instead, a classical output vector can be extracted from a quantum circuit by measuring the expectation values of n locally observable values. We can define this process as the measurement layer (M) that maps quantum states to classical vectors:
\begin{equation}
    \left | y  \right \rangle\to y^{'}=\left\langle y\right| M\left | y  \right \rangle 
\end{equation}

In this work, we design the objective function $L\left ( \cdot  \right )$ to optimize the model parameters by minimizing the error between the real label and the predicted label, improving the classification accuracy. Therefore, we define the optimization objective as:
\begin{equation}
    Loss=minimize\quad  L\left ( y,y^{'}\right )
\end{equation}
When $min$ $L\left ( \cdot  \right )$ is reached, $max\left ( y_{i}^{'}  \right )$ can be obtained for any given PQDs data $X$, then the disturbance type of the data is the ith type.

\subsection{Generation of PQDs and Feature Extraction}
\subsubsection{Mathematical Models of PQDs}
Aiming at the problem that QNN model needs a large amount of data in the training process, this paper obtains the dataset through the PQDs signal mathematical model. According to the International Power Quality IEEE standard and previous studies\cite{ref41}, the PQDs signal mathematical model is used to generate normal power quality signals, seven single types of disturbance signals and three composite types of disturbance signals in the simulation environment. The PQDs signal mathematical model and parameter settings are given as shown in Table 1. The sampling frequency is 1.28 kHz, the fundamental frequency is 50 Hz, the sampling period is 0.2s (10 sampling cycles), and 1000 sample data are generated for each type of signal.
\begin{table*}[t]
\caption{Mathematical model of power quality disturbance signals}\label{Table 1}
\centering
\begin{tabular}{cccc}
\toprule
Label&PQ Disturbance&Mathematical model&Parameter setting \\
\midrule 
$D_{0}$ &normal& \makecell[c]{$V\left ( t \right ) =\sin \left ( \omega t \right )$} & \makecell[c]{$\omega =2\pi f$ ,\\ $f = 50HZ$} \\
$D_{1}$ &harmonic&\makecell[c]{$V(t)=\sin (\omega t)+\alpha_{3} \sin \left(3 \omega t+\varphi_{3}\right)+$ \\ $\alpha_{5} \sin \left(5 \omega t+\varphi_{5}\right)+ \alpha_{7} \sin \left(7 \omega  t+\varphi_{7}\right)$}&\makecell[c]{$0\le \alpha _{3}, \alpha _{5}, \alpha _{7}\le 0.15$, \\ $0\le \varphi _{3},\varphi _{5},\varphi _{7}\le 2\pi$}\\
$D_{2}$&voltage sag&\makecell[c]{$V(t)=\left(1-\alpha\left(u\left(t-t_{1}\right)-u\left(t-t_{2}\right)\right)\right) \sin (\omega t)$}&\makecell[c]{$0.1\le \alpha \le 0.9$, \\ $4T\le t_{2}-t_{1}\le 9T$}\\
$D_{3}$&voltage swell&\makecell[c]{$V(t)=\left(1+\alpha\left(u\left(t-t_{1}\right)-u\left(t-t_{2}\right)\right)\right) \sin (\omega t)$}&\makecell[c]{$0.1\le \alpha \le 0.9$, \\ $4T\le t_{2}-t_{1}\le 9T$}\\
$D_{4}$& voltage interruption&\makecell[c]{$V(t)=\left(1-\alpha\left(u\left(t-t_{1}\right)-u\left(t-t_{2}\right)\right)\right)\sin (\omega t)$}&\makecell[c]{$0.9\le \alpha \le 1$, \\ $4T\le t_{2}-t_{1}\le 9T$}\\
$D_{5}$& voltage flicker&\makecell[c]{$V\left ( t \right ) =\left ( 1+\alpha _{f}\sin\left (\beta \omega t\right)\right)\sin\left (\omega t\right )$}&\makecell[c]{$0.3\le\alpha_{f}\le0.5$, \\ $0.1\le \beta\le0.4 $}\\
$D_{6}$&transient oscillation&\makecell[c]{$V(t)=\sin (\omega t)+\alpha \mathrm{e}^{-\frac{\left(t-t_{3}\right)}{\tau}} \sin \left\{\omega_{n}\left(t-t_{3}\right)\right\}$ \\ $\cdot\left\{u\left(t-t_{3}\right) -u\left(t-t_{4}\right)\right\}$}&\makecell[c]{$0.1\le \alpha\le 0.8$, \\ $0.008\le \tau \le 0.04$ , \\$0.05T\le t_{4}-t_{3}\le 3T$ ,\\$300T\le f_{n}\le 900T$}\\
$D_{7}$&impulsive transient&\makecell[c]{$V(t)=\sin (\omega t)+\alpha \mathrm{e}^{-\frac{\left(t-t_{3}\right)}{\tau}}$ \\ $\left\{u\left(t-t_{3}\right)-u\left(t-t_{4}\right)\right\}$}&\makecell[c]{$1\le \alpha\le 10$ ,\\ $0.008\le \tau \le 0.04$ ,\\ $0.05T\le t_{4}-t_{3}\le 3T$}\\
$D_{8}$&\makecell[c]{voltage sag\\+harmonic}&\makecell[c]{$V(t)=\left(1-\alpha\left(u\left(t-t_{1}\right)-u\left(t-t_{2}\right)\right)\right) \sin (\omega t)$ \\ $+\alpha_{3} \sin \left(3 \omega t+\varphi_{3}\right) +\alpha_{5} \sin \left(5 \omega t+\varphi_{5}\right) $ \\ $ +\alpha_{7} \sin \left(7 \omega t+\varphi_{7}\right)$}&\makecell[c]{$0.1\le \alpha \le0.9$ ,\\ $4T\le t_{2}-t_{1}\le 9T$ ,\\ $0\le \alpha _{3}, \alpha _{5}, \alpha _{7}\le 0.15$ ,\\ $0\le \varphi _{3}, \varphi _{5}, \varphi _{7}\le 2\pi$}\\
$D_{9}$&\makecell[c]{voltage swell\\+harmonic}&\makecell[c]{$V(t)=\left(1+\alpha\left(u\left(t-t_{1}\right)-u\left(t-t_{2}\right)\right)\right) \sin (\omega t)$\\$+\alpha_{3} \sin \left(3 \omega t+\varphi_{3}\right)+\alpha_{5} \sin \left(5 \omega t+\varphi_{5}\right)$\\$+\alpha_{7} \sin \left(7 \omega t+\varphi_{7}\right)$}&\makecell[c]{$0.1\le \alpha \le0.9$,\\$4T\le t_{2}-t_{1}\le 9T$,\\$0\le \alpha _{3},\alpha _{5},\alpha _{7}\le 0.15$,\\ $0\le \varphi _{3},\varphi _{5},\varphi _{7}\le 2\pi$}\\
$D_{10}$&\makecell[c]{voltage interruption\\+harmonic}&\makecell[c]{$
V(t)=\left(1-\alpha\left(u\left(t-t_{1}\right)-u\left(t-t_{2}\right)\right)\right) \sin (\omega t)$ \\ $+ \alpha_{3} \sin \left(3 \omega t+\varphi_{3}\right) +\alpha_{5} \sin \left(5 \omega t+\varphi_{5}\right)$ \\ $+\alpha_{7} \sin \left(7 \omega t+\varphi_{7}\right)$}&\makecell[c]{$0.9\le \alpha \le1$,\\ $4T\le t_{2}-t_{1}\le 9T$,\\ $0\le \alpha _{3},\alpha _{5},\alpha _{7}\le 0.15$ ,\\ $0\le \varphi _{3},\varphi _{5},\varphi _{7}\le 2\pi$}\\
\bottomrule
\end{tabular}
\end{table*}

\subsubsection{S-transform PQDs Analysis and Feature Extraction}
ST is a time-frequency analysis method which is inherited and developed from WT and STFT\cite{ref34}. It was first proposed by Stockwell in 1996. ST can be regarded as a "phase correction" of WT, and can also be evolved from STFT. ST of signal $h\left ( t \right )$ is defined as:
\begin{equation}
    S\left ( \tau ,f \right )=
\int_{-\infty }^{\infty } h\left ( t \right ) \frac{\left | f \right | }{\sqrt{2\pi}}
e^{-\frac{\left (  t-\tau \right ) ^{2} f^{2} }{2} } e^{-j2\pi ft} dt
\end{equation}
where $t$ is the time, $f$ is the frequency, $j$ is the imaginary unit, and $\tau$ is the center position of the Gaussian window function.

Let $\tau \to mT,f\to n/NT$ ($T$ is the sampling period and $N$ is the total number of sampling points), the discrete form of ST can be obtained according to Eq.(12) as:
\begin{equation}
\left\{\begin{matrix}
   S\left [ mT,\frac{n}{NT}\right]= \sum_{k=0}^{N-1}H\left [ \frac{k+n}{NT}\right]e^{-\frac{2\pi^{2}k^{2}}{n}}e^{\frac{j2\pi km}{N}},n\ne 0\\
S\left [ mT,0 \right ] =\frac{1}{N} \sum_{m=0}^{N-1} h\left ( mT \right ),n=0
\end{matrix}\right.
\end{equation}
where $H\left [ k \right ]$  is the discrete Fourier transform of the time series $h\left ( m \right )$, i.e:
\begin{equation}
    H\left [ k \right ] =\frac{1}{N} \sum_{m=1}^{N-1} h\left ( m \right ) e^{-\frac{j2\pi km}{N} } 
\end{equation}
ST of the PQDs signal using Eq.(13) results in a two-dimensional time-frequency complex matrix with frequency-dependent rows and time-dependent columns.

ST performs multi-resolution analysis of time-varying signals, giving high time resolution at high frequencies and high frequency resolution at low frequencies. Since PQDs make power signals non-stationary, ST can be effectively applied. In this paper, seven kinds of single PQDs and many kinds of composite disturbances are simulated by using matlab simulated signals and the features of various disturbances are extracted from the S-matrix. According to the characteristics of various PQDs in time-frequency distributions, the standard statistical techniques are applied to the S-matrix to extract the following nine features $\left ( F_{1},F_{2},\cdots ,F_{9}\right )$\cite{ref25}. Specific characterizations are as follows:

$F_{1}$: The part of the fundamental amplitude time curve greater than 1.02 p.u. accounts for the time proportion of the whole detection time.

$F_{2}$: The part of the fundamental amplitude time curve lesser than 0.98 p.u. accounts for the time proportion of the whole detection time.

$F_{3}$: The part of the fundamental amplitude time curve lesser than 0.15 p.u. accounts for the time proportion of the whole detection time.

$F_{4}$: Sum of lower harmonic skewness. Eq.(15) is used to calculate the skew of the $2\sim 7$ harmonics, and the sum is $F_{4}$.
\begin{equation}
    S=\frac{1}{n}\sum_{i=1}^{n}\left [ \left ( \frac{h_{i}-\mu  }{\sigma }  \right ) ^{3}  \right ]
\end{equation}
where $S$ is the skew of a certain harmonic amplitude sequence, $h_{i}$ is the value of the ith element of the amplitude sequence, $\mu$ is the mean value of the sequence, $\sigma$ is the standard deviation of the sequence and $n$ is the length of the sequence.

$F_{5}$: Sum of middle harmonic kurtosis. Eq.(16) is used to calculate the kurtosis of $8\sim 18$ harmonics respectively, and the sum is obtained to $F_{5}$.
\begin{equation}
    K= \frac{1}{n}\sum_{i=1}^{n}\left [ \left ( \frac{h_{i}-\mu  }{\sigma }  \right ) ^{4}  \right ]
\end{equation}
where $K$ is the kurtosis of a harmonic amplitude sequence.

$F_{6}$: Sum of middle harmonics standard deviations. Eq.(17) is used to calculate the standard deviation of $8\sim 18$ harmonics respectively, and the sum is obtained to $F_{6}$.
\begin{equation}
    SD=\sqrt{\frac{\sum_{i=1}^{n}\left ( h_{i}-\mu \right ) ^{2}}{n-1} }
\end{equation}
where $SD$ is the standard deviation of a harmonic amplitude sequence.

$F_{7}$: Sum of higher harmonic kurtosis. Eq.(16) is used to calculate the kurtosis of $19\sim 30$ harmonics respectively, and the sum is obtained to $F_{7}$.

$F_{8}$: Sum of higher harmonics standard deviations. Eq.(17) is used to calculate the standard deviation of $19\sim 30$ harmonics respectively, and the sum is obtained to $F_{8}$.

$F_{9}$: Average value of total harmonic distortion ($THD$).
\subsection{The Proposed QNN Model}
\begin{figure*}[t]
	\centering
		\includegraphics[width=1.0\linewidth]{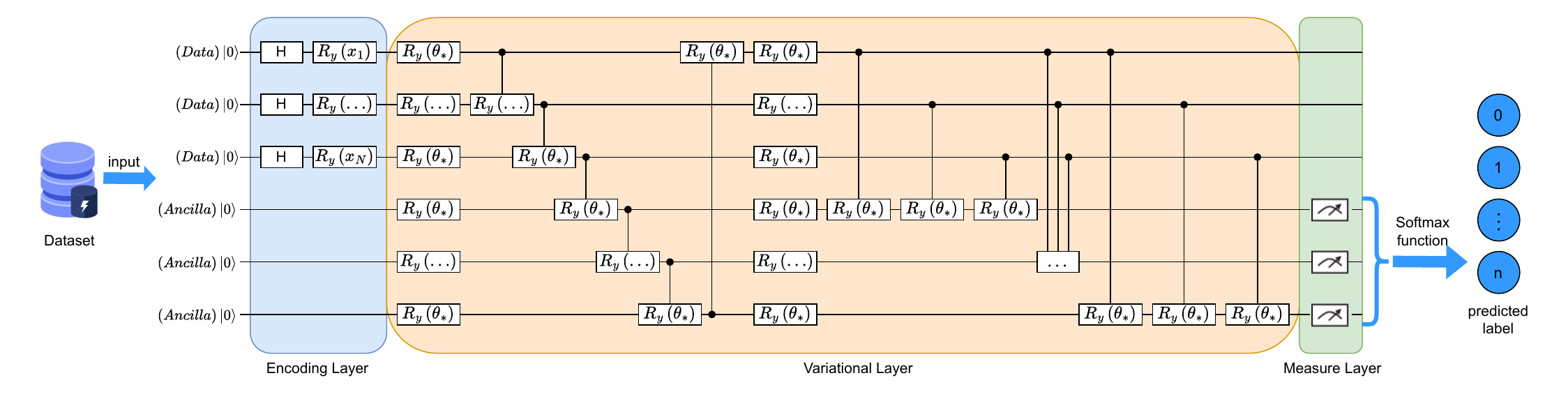}
	\caption{The structure of the proposed quantum neural network model describes the three-layer design in detail.}
	\label{FIG:4}
\end{figure*} 
The structure of the QNN model designed for PQDs recognition is shown in Fig.4. The QNN model consists of three parts: data encoding layer, variational layer and quantum measurement layer. The qubits required for the quantum circuit are based on data qubits and ancilla qubits, where the number of data qubits is equal to the classical data vector dimension, and the number of ancilla qubits is equal to the number of disturbance types (e.g., m disturbance classification recognition requires m ancilla qubits). Initially, all qubits in the quantum circuit are $\left | 0  \right \rangle$. In order to classify PQDs data on the quantum circuit, PQDs data are converted into n-dimensional vectors, which are converted into quantum states through the data encoding layer. The variational layer is designed to set up multi-layer entanglement rotating gates on data qubits and ancilla qubits using $R_{y}$ gates and $CR_{y}$ gates that are adjustable in parameters. Finally, ancilla qubits are measured and the measurement output values are processed to obtain disturbance category labels. Parameters of each layer in the variational layer are adjusted iteratively by using optimization techniques. After adjusting gate parameters, the final circuit can predict the category labels of given PQDs with desired accuracy. 
\subsubsection{Encoding Layer}
Since the PQDs dataset $D=\left (  X^{d},y^{d}  \right )_{d=1}^{D}$ (where $X^{d}=\left [ x_{1}^{d},\dots ,x_{9}^{d} \right ]$ is the input data that can be encoded into a quantum state with $y^{d}$ as the category label for the corresponding data vector) has different dimension between different features and large differences between values. In order to eliminate the possible impact caused by differences in dimension and value range between features, PQDs data are standardized, then the data standardization is calculated as follows:
\begin{equation}
    x_{i}^{*} =\frac{x_{i}-\bar{x}  }{\delta }
\end{equation}
where $x_{i}$ is the ith feature in the PQDs feature vector, $\bar{x}$ is the mean of the original feature, and $\delta$ is the standard deviation of the original feature.

In QNN, the first step is to encode classical information onto quantum states. Most methods of quantum coding can be seen as a parameterized circuit $U\left ( X \right )$ acting on the initializing $\left | 0  \right \rangle ^{\otimes N}$ state, and parameters in the parameterized circuit are determined by the classical information $X$. Different algorithms require different encoding methods. The most efficient approach in terms of space is to encode classical data in the amplitude of the superposition using N qubits to encode $2^{N}$-dimensional data vectors. Another simpler approach is to encode each element of a classical data vector as the amplitude of a single qubit using N qubits to encode an N-dimensional data vector. This encoding method is obviously efficient in terms of time, as it requires only single qubit rotation. Therefore, we choose single qubit amplitude coding to encode PQDs data.

In the QNN model, the initial state of qubits are $\left | 0  \right \rangle$. First, we put qubits in equal superposition states $\left | \phi   \right \rangle=\frac{1}{\sqrt{2} }\left | 0 \right \rangle+\frac{1}{\sqrt{2} }\left | 1  \right \rangle$ of $\left | 0  \right \rangle$ and $\left | 1  \right \rangle$ through Hadamard gates. The superposition state $\left | \phi   \right \rangle$ then transforms the data element $x_{i}^{*}$ in the PQDs data vector through the single qubit rotating gate $R_{y}$ to:
\begin{equation}
    \left | \psi\left ( x_{i}^{*}\right )\right \rangle =R_{y}\left ( \sin^{-1} x_{i}^{*}  \right )\left | \phi   \right \rangle
\end{equation}
All data elements of PQD data vector are encoded in a similar manner. Finally, the encoded data vector is provided to the variational layer as a tensor product, where the quantum states output to the variational layer are
\begin{equation}
    \left | \psi\left ( X \right )   \right \rangle =\otimes _{i=1}^{N} \left | \psi  \left ( x_{i}^{*}  \right ) \right \rangle 
\end{equation}
\subsubsection{Variational Layer}
\begin{figure*}[t]
	\centering
		\includegraphics[width=1.0\linewidth]{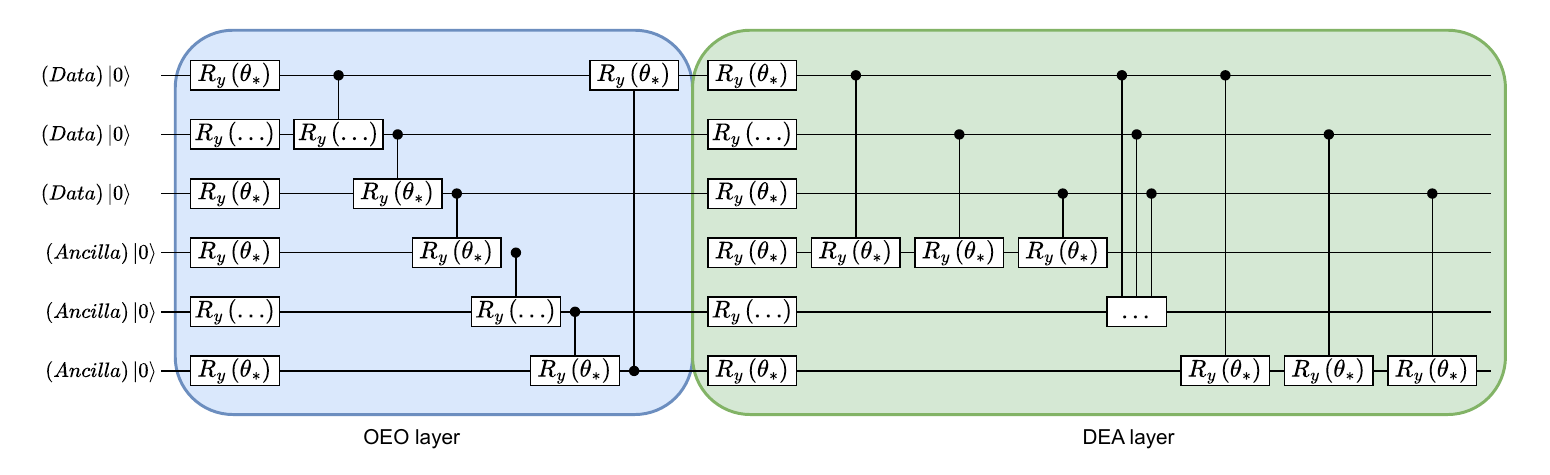}
	\caption{Detailed design of variational layer.}
	\label{FIG:5}
\end{figure*} 
In the proposed QNN framework, PQDs data vectors are embedded with qubits through the data encoding layer. The encoded qubits are then processed through the variational layer and characterized by their trainable parameters, which contribute to rotational operations on qubits. As shown in Fig.5, the variational layer is composed of $OEO$ and $DEA$ layers, where both layers use $R_{y}$ gates and $CR_{y}$ gates to realize quantum transformation. In the $OEO$ layer, first all qubits go through the $R_{y}$ gates to perform rotation operations along the y-axis, such that qubit operations are denoted as
\begin{equation}
    \left | \psi _{i}^{'}   \right \rangle =R_{y}\left ( \theta _{i}  \right ) \left | \psi_{i}\right \rangle 
\end{equation}
where $\left | \psi_{i}\right \rangle$ is the ith qubit state after the initial encoding layer, $\left | \psi _{i}^{'}\right \rangle$ denotes its state after the rotation operation at the $OEO$ layer. The operators $R_{y}\left ( \theta _{i}  \right )$ correspond to rotations at the y-axis with trainable parameters $\theta$. Each qubit is individually manipulated with different rotation parameters to enhance the generalisation of the QNN. Variational operations on each qubit can change the quantum information carried, enabling optimization of subsequent processing steps. However, performing variational operations on single qubit is not sufficient to accurately represent the objective function. Thus, a connection is established between the circuits of each qubit by entangling neighboring qubits through the $CR_{y}$ gate operation. Subsequently, a series of $CR_{y}$ gate operations connect each neighboring qubit, denoted:
\begin{equation}
\begin{split}
    \left | \psi _{out}^{'} \right \rangle = 
    CR_{y} \left ( \theta _{1}, \left | \psi _{1}^{'} \right \rangle, CR_{y}\left ( \theta _{2}, \left | \psi _{2}^{'} \right \rangle, \right. \right. \\ 
    \left. \left. CR_{y} \left ( \dots CR_{y} \left ( \theta _{n-1}, \left | \psi _{n-1}^{'} \right \rangle, \left | \psi _{n}^{'} \right \rangle \right ) \dots \right) \right)  \right) 
\end{split}
\end{equation}
where, $\left | \psi _{out}^{'} \right \rangle$ denotes the final entangled state of the $OEO$ layer and culminates in a $CR_{y}$ operation between $\left | \psi _{out}^{'} \right \rangle$ and $\left | \psi _{1}^{'}   \right \rangle$ resulting into an entanglement that integrates all qubits in $OEO$ layer.

In the $DEA$ layer, $R_{y}$ gates are used to perform rotation operations on qubits that has performed the $OEO$ layer operations to further adjust the state of qubits. The entanglement of data and ancilla qubits are implemented immediately after using $CR_{y}$ gate operations, where data qubits control the rotation of ancilla qubits in a specific way. During the classification and identification process, the input data qubits are entangled with the ancilla qubits to better realize the information interaction, so as to provide the information needed for classification in the subsequent measurement steps. The expressibility and entanglement ability of variational layers have a profound impact on the performance of QNN.

The variational layer in the QNN model is a combined layer of of OEO layer and DEA layer. Typically, a model can include multiple layers that systematically interweave the operations of the variational layer on qubits. After data encoding, the variational layer can be repeated multiple times to increase the model depth and improve the training degrees of freedom to ensure the iterative refinement of the quantum state of PQDs data. Therefore, the input state X of the PQDs data is applied to the multilayer variational layer $U\left ( \theta  \right )$, which can be expressed as:
\begin{equation}
    \left | \psi_{out}   \right \rangle = U\left ( \theta  \right ) \left | \psi\left ( X \right )   \right \rangle 
\end{equation}
\begin{equation}
\begin{split}
    U\left ( \theta  \right ) &=U_{L}\left ( \theta _{L}  \right )\dots U_{l}\left ( \theta _{l}  \right )\dots U_{1}\left ( \theta _{1}  \right )
\\ &=\prod_{l=1}^{L}U_{l}\left ( \theta _{l}  \right )
\end{split}
\end{equation}
where $\left | \psi _{out} \right \rangle$ in Eq.(23) denotes the final output state of the QNN model and defines the overall unitary transformation $U\left ( \theta  \right )$ of the circuit. Where $\theta =\theta _{L},\theta _{L-1},\dots ,\theta _{1}$ denotes the rotation parameter controlling the unitary transformation of the corresponding circuit layer. Eq.(24) is the set of operations for L-layer variational layers.
\subsubsection{Measurement Layer}
Finally, measurements of qubits are needed to extract the information stored in them. Thus, after processing PQDs data vectors using multiple variational layers involving rotation and entanglement, the Pauli-Z expectation values of ancilla qubits are measured. The QNN model measures K ancilla qubits on a computational basis with the expectation value $E\left ( \theta  \right )$ as the output:
\begin{equation}
    E\left ( \theta  \right ) =\left\langle\psi\left ( X \right ) \right| U^{\dagger }\left ( \theta  \right ) Z^{\otimes K}U\left ( \theta  \right )\left | \psi \left ( X \right )  \right \rangle    
\end{equation}

Now, the expected values of Pauli-Z from the measurement of ancilla qubits are passed through the $softmax$ function. The input values of the $softmax$ function (S) are converted to a set of probabilities, i.e:
\begin{equation}
    S\left ( \left\langle\psi _{i} \right| Z\left | \psi_{i}  \right \rangle  \right )
=\frac{exp\left ( \left\langle\psi_{i}\right| Z\left | \psi_{i}  \right \rangle  \right ) }
{\sum_{i=1}^{K}exp\left ( \left\langle\psi_{i}\right| Z\left | \psi_{i}  \right \rangle  \right ) }  
\end{equation}
where $\left\langle\psi_{i}\right|Z\left | \psi_{i}  \right \rangle$ gives the Pauli-Z measurement on ancilla qubit denoted by $\left | \psi_{i}  \right \rangle$. The category corresponding to the maximum probability obtained by Eq.(26) is the classification label of the disturbed signal.
\subsubsection{Model Training Learning Process}
Parameter optimization and updating is an important factor to determine the effectiveness of the model, so the model learning and training process is crucial. In quantum-classical hybrid models, backpropagation methods based on gradient descent are usually used to optimize the parameters in the learning process. In our model, the optimization technique based on gradient descent is also used to update parameters so that the objective function converges to the minimum value.

For given training samples, we calculate the loss $L$ using the objective function $\ell\left ( \cdot  \right )$:
\begin{equation}
    L=\ell \left ( y,E\left ( \theta  \right )  \right )
\end{equation}

Due to slightly different problem solving, different loss functions are used. Throughout the training process, the binary cross-entropy loss function and the categorical cross-entropy loss function are used for PQDs detection binary classification and PQDs recognition multi-classification, respectively, in order to quantify difference between the true category labels and the predicted category probabilities, with the overall goal of minimizing the difference. The binary cross-entropy loss function is given by the following equation:
\begin{equation}
\begin{split}
     L_{BCE}\left ( y,E\left ( \theta  \right ) \right ) 
     =-\frac{1}{n} \sum_{i=1}^{n} \left[ y_{i} \cdot log \left ( E \left ( \theta \right )   \right ) \right. \\ \left.
     +\left ( 1-y_{i} \right )\cdot log\left ( 1-E\left ( \theta  \right ) \right ) \right]
\end{split}  
\end{equation}
In Eq.(28), $y_{i}$ is the actual label, denoting the label 0 or 1 corresponding to normal and abnormal signals, and $n$ is the number of samples. The categorical cross-entropy loss function is given by the following equation:
\begin{equation}
     L_{CCE}\left ( y,E\left ( \theta  \right ) \right ) =-\frac{1}{n}\sum_{i=1}^{n}\sum_{j=1}^{m} y_{ij}log\left ( E\left ( \theta  \right )  \right )   
\end{equation}
In Eq.(29), $y_{ij}$ is the $one-hot$ code representing the true label of PQDs, $n$ is the number of samples, and $m$ is the number of categories.

We use the analytic gradient\cite{ref42}, which is best suited for gate parameter updating, to avoid the effects of noise. The analytic gradient is calculated as the derivative of the output value $\theta$  of the input. Therefore, the partial derivative of the loss function is given by the following:
\begin{equation}
    \frac{\partial L}{\partial \theta _{i} } =\frac{\partial L}{\partial E}.\frac{\partial E}{\partial \theta _{i} }  
\end{equation}
$\frac{\partial L}{\partial E}$ is easily obtained from the objective function $\ell\left ( \cdot  \right )$, so only $\frac{\partial E}{\partial \theta _{i} }$ needs to be calculated.

We use the parameter shift rule\cite{ref35} to calculate the gradient of the expectation value of an observable with respect to the circuit parameter $\theta$, where the partial derivatives of the quantum circuit are given by $E\left ( \theta  \right )$:
\begin{equation}
\begin{split}
\frac { \partial E \left ( \theta \right ) } { \partial \theta _{i} } =
\frac{1}{2} \left ( \left \langle \psi \left ( X \right ) \right | U^{\dagger} \left ( \theta _{i}^{+} \right ) Z^{ \otimes K} U \left ( \theta _{i}^{+} \right ) \left | \psi \left ( X \right ) \right \rangle \right. \\ \left. \left \langle \psi \left ( X \right ) \right | U^{ \dagger } \left ( \theta _{i}^{-} \right ) Z^{ \otimes K } U \left ( \theta _{i}^{-} \right ) \left | \psi \left ( X \right ) \right \rangle \right )
\end{split}    
\end{equation}
where $\theta _{i}^{\pm }$ denotes the operation of $\pm \frac{\pi }{2}$ on the ith parameter in $\theta$.

At each iteration $t$, we estimate the gradient by setting the learning rate $\eta$. Set update parameter $\theta _{i}$ as follows:
\begin{equation}
    \theta _{i}^{t+1}\gets \theta _{i}^{t}-\eta \frac{\partial L\left ( \theta  \right ) }{\partial \theta _{i}^{t} }  
\end{equation}
Instead of computing on the entire training set, this method computes on small batches of data, thus avoiding optimization of local minima. Using the gradient descent method, the parameters are iteratively updated to ensure that the QNN model adapts to and learns the PQDs classification. After the training is completed, the QNN-based PQDs classification model can obtain the type probability value of each disturbance sample through through the output of QNN. The specific model training learning process can be seen in Fig.6. We summarize the proposed QNN algorithm for solving PQDs classification and recognition in Algorithm 1.
\begin{algorithm}[H]
    \caption{The proposed QNN algorithm}
    \begin{algorithmic}[1]
        \Require A PQDs classical dataset $D = \left ( X^{d}, y^{d} \right ) _{d=1}^{D}$, $X^{d} \in R^{N}$
        \Ensure A classification result $y \in \left \{ 0,1 \dots n \right \}$
        \State  Initialize $:$ Data qubits $\gets \left | 0 \right \rangle$, Ancilla qubits $\gets \left | 0 \right \rangle$
        \State m $\gets$ number of layers
        \State $\theta \gets$ quantum gate parameters optimized using classical computer
        \State Data qubits $\gets$ Data Encoder$\left ( X^{d} \right )$
        \For {i $\gets$ 1 to m} 
             \State $R_{y}\left ( \theta \right )$ rotations on Data qubits and Ancilla qubits
             \State $CR_{y}\left ( \theta \right )$ entangles each neighboring qubit
             \State $R_{y}\left ( \theta \right )$ rotations on Data qubits and Ancilla qubits
             \State $CR_{y}$ ( $\theta$, Data qubits, Ancilla qubits)
        \EndFor
        \State $E\left ( \theta \right )\gets$ measurement ( Ancilla qubits )
        \State class label $\gets softmax\left ( E\left ( \theta \right )  \right )$
        \If {training}
           \State (1) Maximize the classification accuracy of the training dataset using  cross-entropy.
           \State (2) Optimizing parameters using gradient descent, namely $\theta _{i}^{t+1}\gets \theta _{i}^{t}-\eta \frac{\partial L\left ( \theta \right ) }{\partial \theta _{i}^{t} }$ back to the quantum circuit.
        \Else \If{testing}
        \State Evaluate the performance by checking the cost function concerning a dataset taken independently from the training one.
        \EndIf
       \EndIf
    \end{algorithmic}
\end{algorithm}
\begin{figure}[h]
		\centering
		\includegraphics[width=1\linewidth]{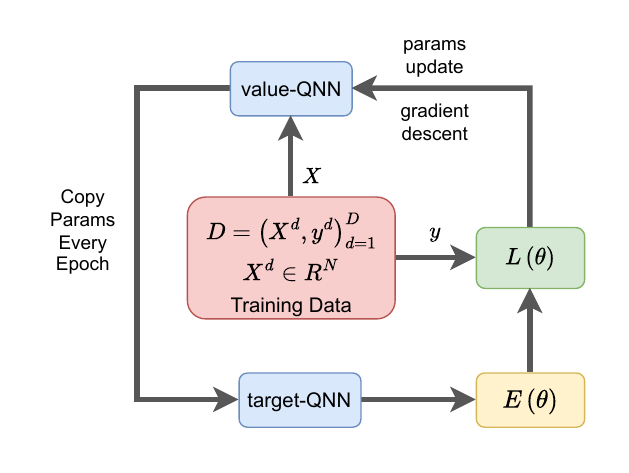}
		\caption{Model training learning process.}
		\label{FIG:6}
\end{figure}
\subsubsection{Complexity Analysis}
The metrics to quantify the expressive power of algorithms are the runtime complexity and the space complexity to simulate certain probability distributions. According to conventions\cite{ref45}, the runtime complexity of algorithms based on PQC refers to the gate complexity\cite{ref46}.

In order to analyse the complexity of the proposed QNN algorithm, the circuit is extended to $L$ blocks, where each block applies $U_{l}\left ( \theta _{l}  \right )$, with $l\in \left [ 1,L \right ]$ and $L\sim poly\left ( N \right )$. The structure is consistent with the variational layer of Fig.4 in each block. The circuit depths $L$ grows only polynomially with the number of qubits $N$, namely, the runtime complexity of the QNN algorithm is $O\left ( poly\left ( N \right )  \right )$. The space complexity depends on the encoding. In our scheme, single qubit amplitude encoding is used. However, the number of qubits required for QNN is determined by data qubits and ancilla qubits. Where data qubits are used for data encoding. Assuming a $n$-class PQDs classification recognition problem using $n$ data qubits for data encoding and $m$ ancilla qubits, the space complexity of the QNN algorithm is $O\left ( N \right ) = O\left ( m+n \right )$ at this point, where $N$ denotes the number of qubits required to construct the quantum circuit.
\section{Experiment}
In this section, we design experiments to evaluate and validate the proposed QNN-based detection and recognition of PQDs. The next section describes in detail the experimental configuration, the generation of experimental datasets, and the analysis of a series of experimental results.
\subsection{Experimental Configuration}
To verify the feasibility of the models in this paper for experiments, the cross-platform software library Mindquantum is used to construct and measure the quantum circuits designed. The traditional model partially built using the Mindspore framework. We use Adam optimizer,which is computationally cheaper and easier to implement than other optimizers such as stochastic gradient descent-based optimizer. The experimental environment is Windows 11 Professional, Intel(R) Core(TM) i7-8550U CPU @1.80GHz, and NVIDIA GeForce GTX 1650 8 GB GPU. The PQDs dataset is generated from the PQDs mathematical model using matlab2020.
\subsection{Experimental Dataset}
The data required during the training of the QNN model in this paper is generated in the matlab simulation environment based on the PQDs mathematical model and feature extraction method in Section 3.2. In total, we generate 11 types of PQDs signal data, each of which containe 1000 entries. The dataset is divided according to the needs of subsequent experiments.
\subsection{Experimental Results and Analysis}
\subsubsection{PQDs Detection Two Classifications}
It is also significant to detect whether there are disturbances in the power system before recognizing the specific type of PQDs. In this experiment, we mainly detect whether there are PQDs and verify whether the model can screen out the disturbance data. The dataset used to detect PQDs contains 2000 data, of which 1000 data are normal signals and the remaining 1000 data are random PQDs signals. The quantum circuit structure for this experiment is shown in Fig.4, where only two ancilla qubits are required. During the experiment, batch size is set to 32, and the training last 25 epochs. As shown in Fig.7, the classification accuracy gradually increases with the increase of the training epochs, and the highest training accuracy and test accuracy reach 100\% and 99.75\% respectively. As shown in Fig.8, the loss rate declined rapidly during the first 15 epochs, and then gradually reached convergence. The experiment proves that the designed QNN model can screen out the disturbance signal accurately and achieve the purpose of detection.
\begin{figure}[h]
		\centering
		\includegraphics[width=1\linewidth]{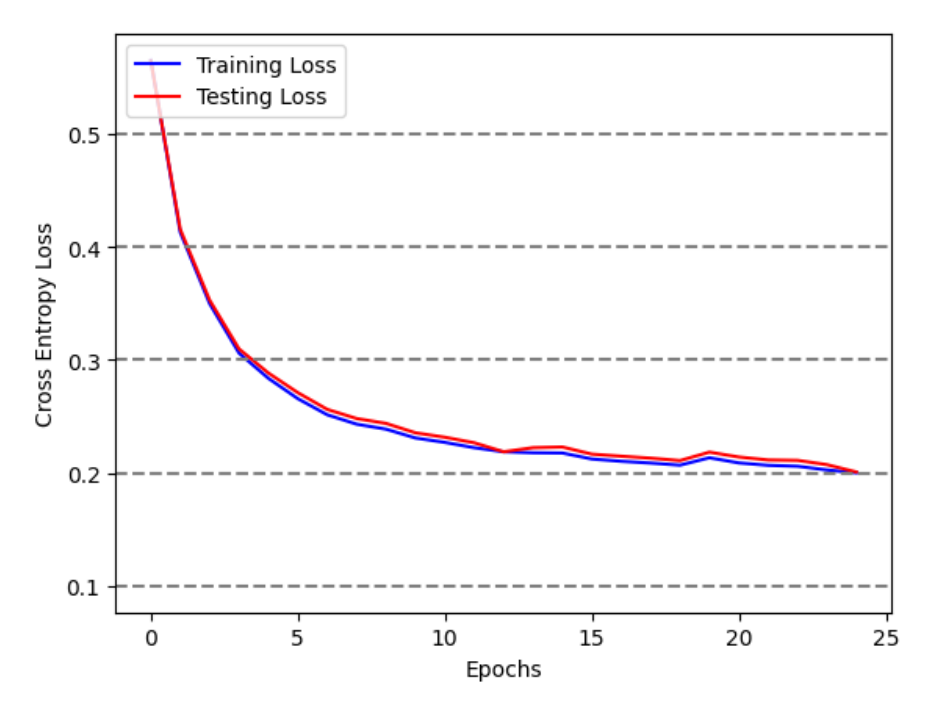}
		\caption{ PQDs detection binary classification loss rate.}
		\label{FIG:7}
\end{figure}
\begin{figure}[h]
		\centering
		\includegraphics[width=1\linewidth]{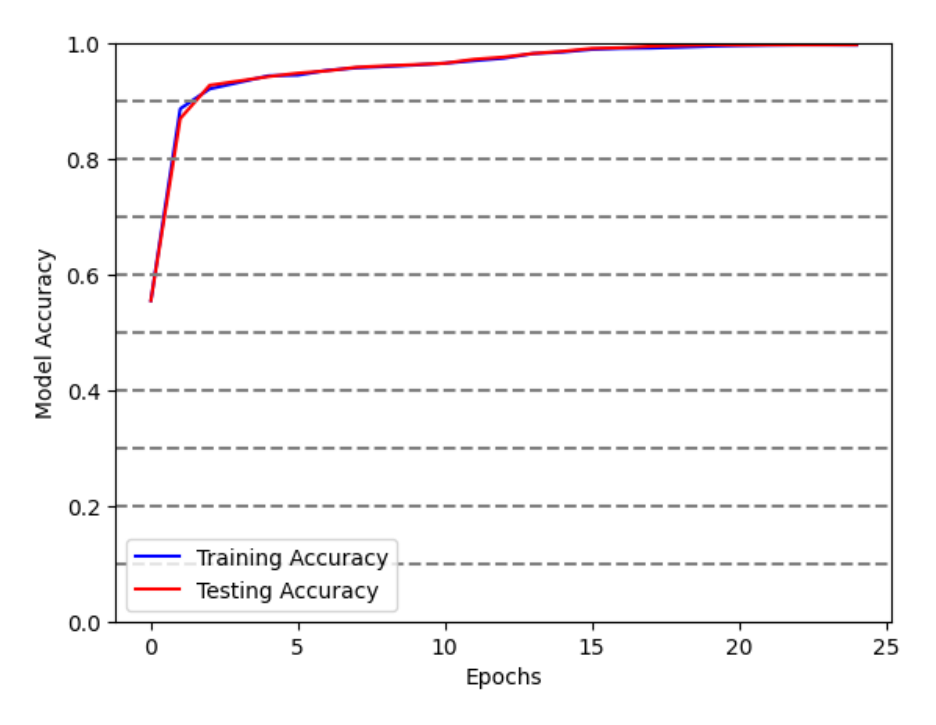}
		\caption{ PQDs detection binary classification accuracy.}
		\label{FIG:8}
\end{figure}
\subsubsection{PQDs Recognition Multi-Classification}
Since PQDs exist as single disturbances and composite disturbances, PQDs identification usually involves PQDs multi-classification. The quantum circuit of the QNN model can adjust the number of ancilla qubits according to the problem solved, and expand the required quantum circuit.First, we perform a classification experiment to recognize seven single PQDs. During the experiment, batch size is set to 16, and the training last 105 epochs. As shown in Fig.9, the classification accuracy increases rapidly in the first 30 epochs, followed by gradual convergence. The best training accuracy and test accuracy of the QNN model on a single PQDs dataset are 98.39\% and 97.85\%, respectively.
\begin{figure}[h]
		\centering
		\includegraphics[width=1\linewidth]{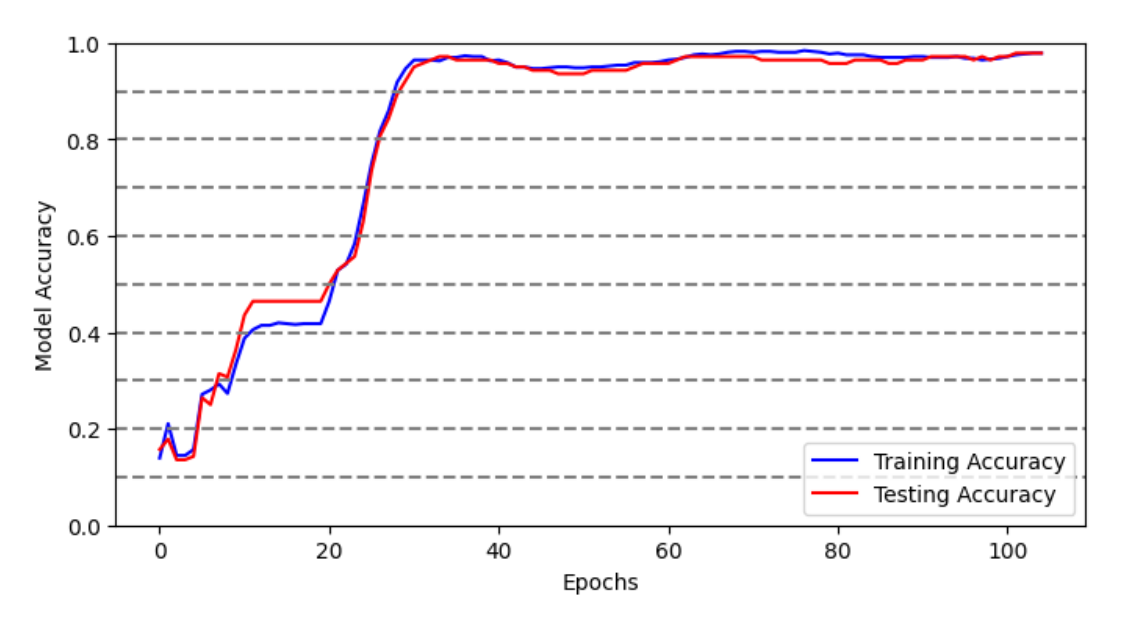}
		\caption{ PQDs seven classification accuracy.}
		\label{FIG:9}
\end{figure}
The confusion matrix shown in Fig.10 describes the number of correct and incorrect predictions of single PQDs classes. It can be observed that there is a small error in the prediction results obtained by the QNN model for each PQDs class, but the number of correct categories is far greater than the number of wrong categories on the whole. The feasibility of recognizing single PQDs by the QNN model is proved.
\begin{figure}[h]
		\centering
		\includegraphics[width=1\linewidth]{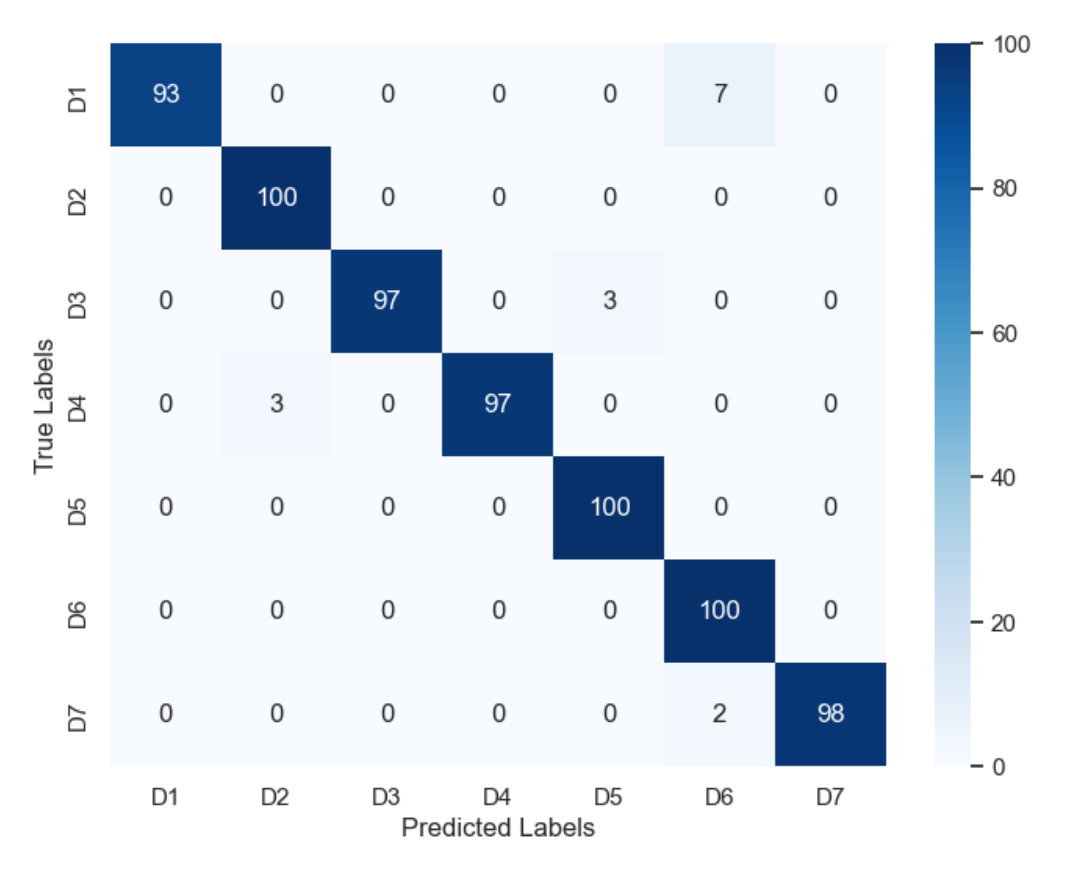}
		\caption{ PQDs seven classification confusion matrix.}
		\label{FIG:10}
\end{figure}

With the increasing complexity of power system access signals, PQDs often appear in the form of mixed PQDs rather than traditional single disturbances. Therefore, in order to increase the universality, we perform classification experiments recognizing mixed PQDs, i.e., containing seven single PQDs signals and three composite PQDs signals. The batch size is set to 10 for the experiment, and 160 epochs of training are performed. As shown in Fig.11, the accuracy gradually increases with the increase of training epochs, with the highest training accuracy and test accuracy reaching 95.625\% and 95.5\%, respectively. The confusion matrix shown in Fig.12 describes the number of correct and incorrect predictions of mixed PQDs classes. It can be observed that $D_{6} \sim D_{10} $ has a larger classification error than other classes, which may be caused by similar waveforms. It is proved that the QNN model can be extended to recognize mixed disturbances.
\begin{figure}[h]
		\centering
		\includegraphics[width=1\linewidth]{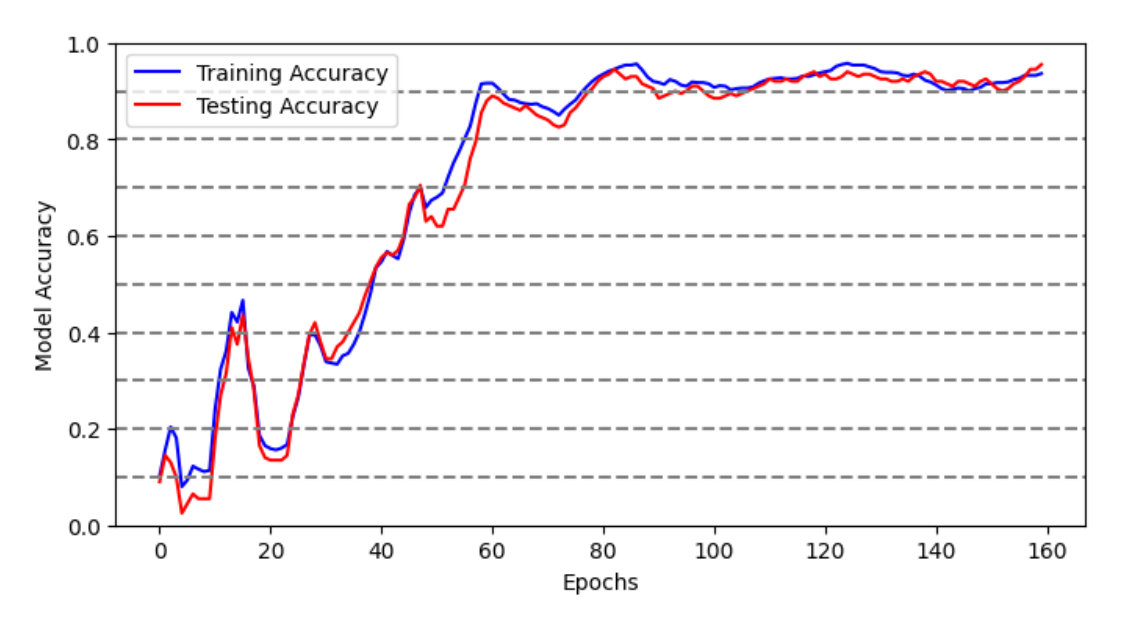}
		\caption{ PQDs ten classification accuracy.}
		\label{FIG:11}
\end{figure}
\begin{figure}[h]
		\centering
		\includegraphics[width=1\linewidth]{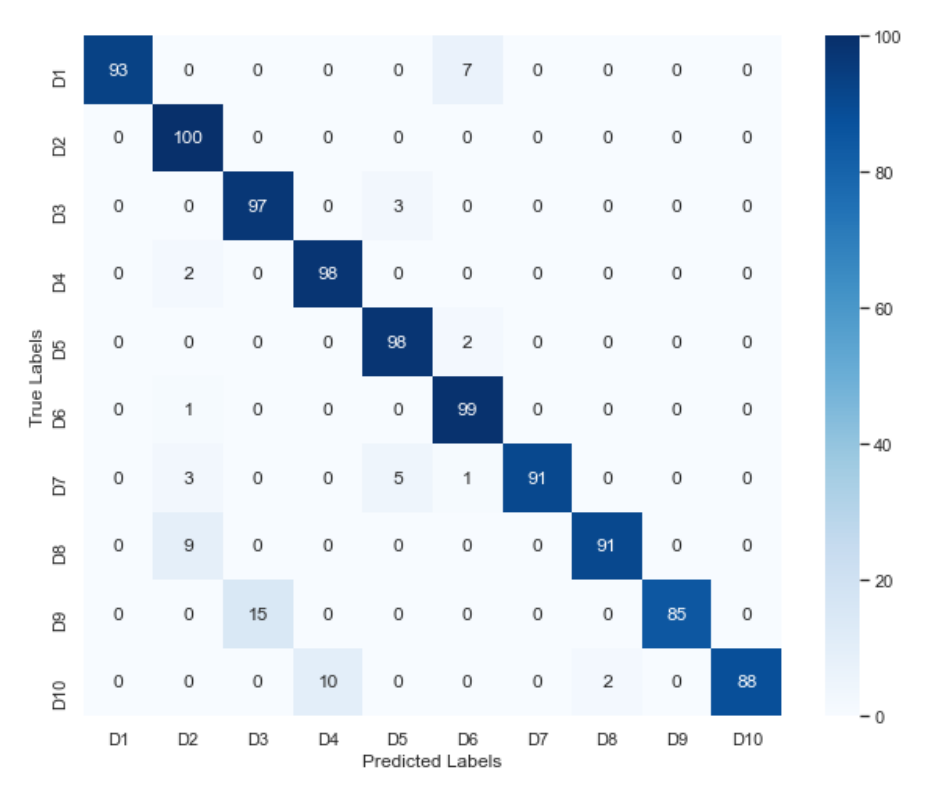}
		\caption{ PQDs ten classification confusion matrix.}
		\label{FIG:12}
\end{figure}

\subsubsection{PQDs Recognition under Different Noise Conditions}
The PQDs dataset in this study is generated according to the methodology outlined in 3.2. However, real PQDs signals collected from power systems are often accompanied by noise. To explore the effects of the proposed QNN in different levels of noise, we introduce Gaussian white noise with signal-to-noise ratios(SNR) of 20dB, 30dB, and 50dB into seven single PQDs signals. As shown in Table 2, the accuracy decreases by 1\%, 0.28\%, and 0.14\% for noise conditions with SNR of 20 dB, 30 dB, and 50 dB, but the overall accuracies still achieve more than 96\%. As mentioned earlier, the smaller SNR, the less tolerant the QNN model is to noise. These experimental results verify the robustness of the improved QNN model. The model has good anti-noise performance and can still achieve high classification accuracy under noise conditions.
\begin{table}[H]
\caption{Model accuracy under various noise conditions}
\centering
\begin{tabular}{ccccc}
\toprule
Label& \multicolumn{4}{c}{Accuracy(\%) under SNR} \\
&pure&40dB&30dB&20dB\\
\midrule
$D_{1}$&93&93&93&92 \\
$D_{2}$&100&100&100&100 \\
$D_{3}$&97&97&97&98 \\
$D_{4}$&97&97&97&94 \\
$D_{5}$&100&99&99&98 \\
$D_{6}$&100&100&100&96 \\
$D_{7}$&98&98&97&100 \\
Average accuracy&97.85&97.71&97.57&96.85 \\
\bottomrule
\end{tabular}
\end{table}

\subsubsection{Comparative Experiments with the rest of models}
\begin{table*}[!t]
\caption{Comparison with existing methods}\label{Table 3}
\centering
\begin{tabular}{ccccc}
\toprule
Method & Num.of PQDs & Num.of Features & Num.of Parameters & Accuracy\\
\midrule
LeNet-5\cite{ref40}&64&-&6610&93.01\%\\
AlexNet\cite{ref40}&64&-&60000000&90.10\%\\
Net-256\cite{ref40}&64&-&28352&96.05\%\\
STFT+LSTM\cite{ref36}&6&-&-&79.14\%\\
WT+PSO-ELM\cite{ref37}&10&6&-&97.6\%\\
ADALINE+FFNN\cite{ref38}&8&10&-&90\%\\
EWT+SVM\cite{ref39}&9&-&-&96.44\%\\
ST+DT\cite{ref43}&9&-&-&97.37\%\\
WVD+CNN\cite{ref44}&9&-&-&94.44\%\\
QNN(Our Paper)&7&9&120&97.85\%\\
\midrule
\end{tabular}
\end{table*}
To further evaluate the effectiveness of the designed QNN model, we compared it with classical models such as SVM, PNN, FFML, CNN, and LSTM. In order to unify the standard, the same datasets are used and experiments are conducted under the same settings. As can be seen in Fig.14 and Fig.15, the QNN model achieves higher accuracy using fewer parameters compared to other classical models evaluated in the experiment. Although the accuracy of the QNN model to recognize PQDs is similar to some classical models, the number of parameters required is greatly reduced. The experimental results show that the QNN model has some advantages over the classical network in recognizing PQDs.

In order to validate the proposed method, a comparison is made with existing conventional methods. The performance of the different methods under complex PQDs is shown in Table 3. The following results show that the accuracy of the conventional methods for PQDs classification based on feature extraction is slightly lower than that of QNN, while the number of trainable parameters involved in model training by the conventional methods for model training with automatic extraction of features during model training is much more than that of QNN. QNN shows its advantages in different aspects, highlighting its use for solving PQDs recognition.
\begin{figure}[h]
		\centering
		\includegraphics[width=1\linewidth]{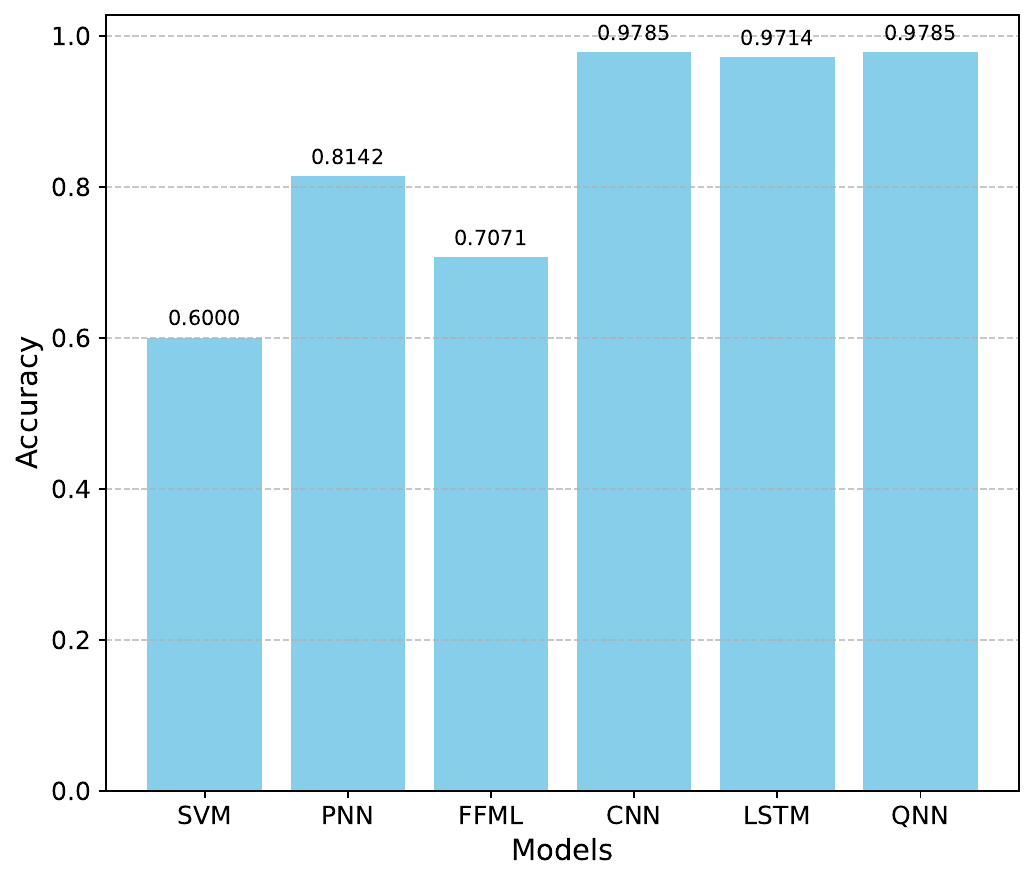}
		\caption{ The accuracy of various models under the same conditions.}
		\label{FIG:13}
\end{figure}
\begin{figure}[h]
		\centering
		\includegraphics[width=1\linewidth]{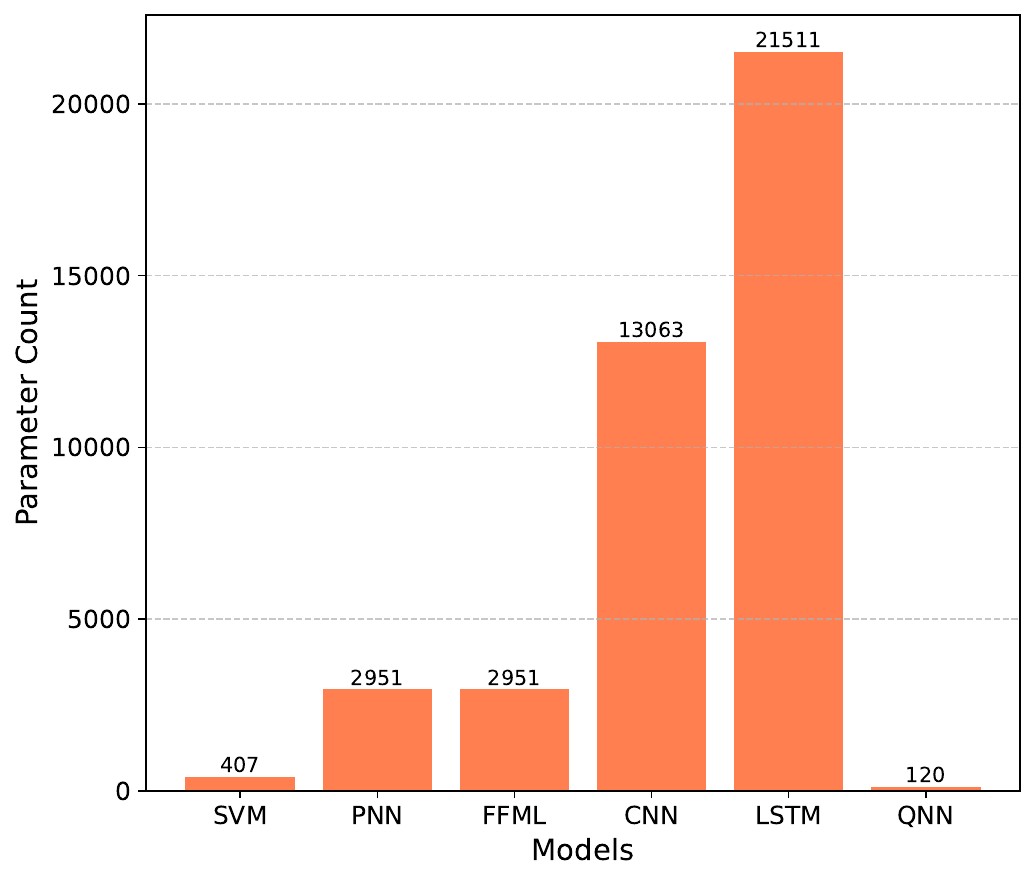}
		\caption{ The parameter count of various models under the same conditions.}
		\label{FIG:14}
\end{figure}

\section{Conclusion}
In this paper, we propose an improved QNN model for the detection and recognition of PQDs. The model leverages the properties of quantum computing to accelerate problem-solving in power quality issues and enhance the efficiency of PQDs detection and recognition. The QNN model demonstrates significantly lower runtime and space complexities, $O\left ( poly\left ( N \right )  \right )$ and  $O\left ( N \right )$ respectively, compared to classical neural networks. The model is designed to extend the quantum circuit for various disturbances detection and recognition tasks. Experimental results show that the proposed QNN model achieves high accuracies of 99.75\% for binary classification in detecting PQDs, 97.85\% for recognizing seven types of single PQDs, and 95.5\% for recognizing ten types of mixed PQDs. Additionally, the QNN model displays strong robustness in PQDs recognition and outperforms classical models in this field. The proposed QNN model presents a promising solution for power quality issues, enhancing the stability and reliability of the power system by improving computational efficiency and accuracy. Although the application of QNN to PQDs detection and recognition is a preliminary exploration, our future research will focus on expanding the model's capabilities and applications further.


\section{Acknowledgments}

This research was funded by the Guizhou Power Grid Co., Ltd. Technology Project (Grant No. GZKJXM20222149).

\bibliographystyle{plain}

\begin{thebibliography}{40}
\bibitem {ref1} R. S. Shaik, R. Kumar, M. Asif and D. Yadav."DTCWT-SVM Based Identification of Single and Multiple Power Quality Disturbances." \href{https://doi.org/10.1109/ICICCSP53532.2022.9862338}{2022 International Conference on Intelligent Controller and Computing for Smart Power (ICICCSP). IEEE, 01-06(2022).}

\bibitem {ref2} Qiu, Wei and Tang, Qiu and Liu, Jie and Yao, Wenxuan."An automatic identification framework for complex power quality disturbances based on multifusion convolutional neural network." \href{https://doi.org/10.1109/TII.2019.2920689}{IEEE Transactions on Industrial Informatics \textbf{16(5)}, 3233-3241 (2020).}

\bibitem {ref3}Fu, Qiang and Montoya, Luis F. and Solanki, Ashish and Nasiri, Adel and Bhavaraju, Vijay and Abdallah, T. and Yu, David C."Microgrid generation capacity design with renewables and energy storage addressing power quality and surety." \href{https://doi.org/10.1109/TSG.2012.2223245}{IEEE Transactions on Smart Grid \textbf{3(4)}, 2019-2027(2012).}

\bibitem {ref4} Liang, Xiaodong. "Emerging Power Quality Challenges Due to Integration of Renewable Energy Sources." \href{https://doi.org/10.1109/TIA.2016.2626253}{IEEE Transactions on Industry Applications\textbf{53(2)}, 855-866 (2017).}

\bibitem {ref5} Lee, Chun-Yao and Shen, Yi-Xing. "Optimal feature selection for power-quality disturbances classification."  \href{https://doi.org/10.1109/TPWRD.2011.2149547}{IEEE Transactions on Power Delivery \textbf{26(4)},2342-2351 (2011).}

\bibitem {ref6} Anggriawan, Dimas Okky and Wahjono, Endro and Sudiharto, Indhana and Firdaus, Aji Akbar and Novita Nurmala Putri, Dianing and Budikarso, Anang. "Identification of short duration voltage variations based on short time Fourier transform and artificial neural network."  \href{https://doi.org/10.1109/IES50839.2020.9231815}{2020 international electronics symposium (IES). IEEE, 43-47(2020).}

\bibitem {ref7} Liu, Ke and Chen, Laijun and Li, Nanfang and Yang, Jian and Han, Jun and Su, Xiaoling."Wavelet Change and Convolutional Neural Network Based Power Quality Online State Estimation Method."  \href{https://doi.org/10.1109/EEI59236.2023.10212475}{2023 5th International Conference on Electronic Engineering and Informatics (EEI). IEEE, 556-561(2023).}

\bibitem {ref8} Mahela, Om Prakash and Khan, Baseem and Alhelou, Hassan Haes and Siano, Pierluigi. "Power Quality Assessment and Event Detection in Distribution Network With Wind Energy Penetration Using Stockwell Transform and Fuzzy Clustering." \href{https://doi.org/10.1109/TII.2020.2971709}{IEEE Transactions on Industrial Informatics \textbf{16(11)},6922-6932(2020).}

\bibitem {ref9} Dash, P.K. and Panigrahi, B.K. and Panda, G."Power quality analysis using S-transform."\href{https://doi.org/10.1109/TPWRD.2003.809616 }{IEEE Transactions on Power Delivery \textbf{18(2)}, 406-411(2003).}

\bibitem {ref10} Zhong, Tie and Zhang, Shuo and Cai, Guowei and Li, Yue and Yang, Baojun and Chen, Yun. "Power quality disturbance recognition based on multiresolution S-transform and decision tree."  \href{https://doi.org/10.1109/ACCESS.2019.2924918}{IEEE Access \textbf{7}, 88380-88392 (2019).}

\bibitem {ref11} Wenjing, Zhao and Liqun, Shang and Sike, Dong and Jinfan, Sun."Power Quality Composite Disturbance Recognition Based on Grid Optimization SVM."  \href{https://doi.org/10.1109/ICECE48499.2019.9058548}{2019 IEEE 2nd International Conference on Electronics and Communication Engineering (ICECE). IEEE, 123-127(2019).}

\bibitem {ref12} Minh Khoa, Ngo and Van Dai, Le."Detection and classification of power quality disturbances in power system using modified-combination between the stockwell transform and decision tree methods."  \href{ https://doi.org/10.3390/en13143623}{Energies\textbf{13(14)}, 3623(2020).}

\bibitem {ref13} Tang, Qiu and Qiu, Wei and Zhou, Yicong. "Classification of Complex Power Quality Disturbances Using Optimized S-Transform and Kernel SVM."  \href{https://doi.org/10.1109/TIE.2019.2952823}{IEEE Transactions on Industrial Electronics \textbf{67(11)}, 9715-9723 (2020).}

\bibitem {ref14} Fu, Lei and Deng, Xi and Chai, Haoqi and Ma, Zepeng and Xu, Fang and Zhu, Tiantian."PQEventCog: Classification of power quality disturbances based on optimized S-transform and CNNs with noisy labeled datasets."  \href{https://doi.org/10.1016/j.epsr.2023.109369}{Electric Power Systems Research\textbf{220},109369(2023).}

\bibitem {ref15} Caro, Matthias C and Huang, Hsin-Yuan and Cerezo, Marco and Sharma, Kunal and Sornborger, Andrew and Cincio, Lukasz and Coles, Patrick J."Generalization in quantum machine learning from few training data." \href{https://doi.org/10.1038/s41467-022-32550-3}{Nature communications \textbf{13(1)}, 4919(2022).}

\bibitem {ref16} BIAMONTE J, WITTEK P, PANCOTTI NBiamonte, Jacob and Wittek, Peter and Pancotti, Nicola and Rebentrost, Patrick and Wiebe, Nathan and Lloyd, Seth." Quantum machine learning." \href{https://doi.org/10.1038/nature23474}{Nature \textbf{549(7671)}, 195-202(2017).}

\bibitem {ref17}Arute, Frank, et al."Quantum supremacy using a programmable superconducting processor." \href{https://doi.org/10.1038/s41586-019-1666-5
}{Nature \textbf{574(7779)},505-510 (2019).}

\bibitem {ref18} Zhong, Han-Sen, et al. "Quantum computational advantage using photons." \href{https://doi.org/10.1126/science.abe8770}{Science\textbf{370(6523)},1460-1463(2020).}

\bibitem {ref19}Kak, Subhash C."Quantum neural computing." \href{https://doi.org/10.1016/S1076-5670(08)70147-2}{Advances in imaging and electron physics \textbf{94}, 259-313 (1995).}

\bibitem {ref20} Abbas, Amira, et al."The power of quantum neural networks." \href{https://doi.org/10.1038/s43588-021-00084-1}{Nature Computational Science \textbf{1(6)}, 403-409 (2021).}

\bibitem {ref21} Abubakar, Muhammad and Nagra, Arfan Ali and Faheem, Muhammad and Mudassar, Muhammad and Sohail, Muhammad.  "High-Precision Identification of Power Quality Disturbances Based on Discrete Orthogonal S-Transforms and Compressed Neural Network Methods."\href{https://doi.org/10.1109/ACCESS.2023.3304375}{IEEE Access\textbf{11},85571-85588(2023).}

\bibitem {ref22} Yu, Chao-Hua and Gao, Fei and Wen, Qiao-Yan." An improved quantum algorithm for ridge regression." \href{https://doi.org/10.1109/TKDE.2019.2937491}{IEEE Transactions on Knowledge and Data Engineering\textbf{33(3)}, 858-866(2019).}

\bibitem {ref23} Lü, Yanxuan and Gao, Qing and Lü, Jinhu and Ogorzałek, Maciej and Zheng, Jin."A Quantum Convolutional Neural Network for Image Classification." \href{https://doi.org/10.23919/CCC52363.2021.9550027}{2021 40th Chinese Control Conference (CCC), 6329-6334 (2021).}

\bibitem {ref24} Zheng, Jin and Gao, Qing and Lü, Yanxuan."Quantum Graph Convolutional Neural Networks."  \href{https://doi.org/10.23919/CCC52363.2021.9550372}{2021 40th Chinese Control Conference (CCC), 6335-6340 (2021).}

\bibitem {ref25} Li, Z.-M., Lv, Q.-Y., Chen, N., Pei, Z.-Y., Ding, Y.-H., and Gong, Y."Power quality composite disturbance identification based on chaotic ensemble decision tree." {Power system protection and control \textbf{49},18–27(2021).}

\bibitem {ref26} L{\"u}, Yanxuan and Gao, Qing and L{\"u}, Jinhu and Pan, Y and Dong, D."Recent advances of quantum neural networks on the near term quantum processor."  \href{https://doi.org/10.1360/SST-2020-0459}{Scientia Sinica Technologica \textbf{52(4)}, 547-564 (2022).}

\bibitem {ref27} da Silva, Adenilton Jos{\'e} and Ludermir, Teresa Bernarda and de Oliveira, Wilson Rosa."Quantum perceptron over a field and neural network architecture selection in a quantum computer." \href{	https://doi.org/10.1016/j.neunet.2016.01.002}{Neural Networks \textbf{76}, 55-64 (2016).}

\bibitem {ref28} Jahn, Alexander and Zimbor{\'a}s, Zolt{\'a}n and Eisert, Jens."Tensor network models of AdS/qCFT." \href{	https://doi.org/10.22331/q-2022-02-03-643}{Quantum \textbf{6}, 643 (2022).}

\bibitem {ref29} Hur, Tak and Kim, Leeseok and Park, Daniel K."Quantum convolutional neural network for classical data classification." \href{https://doi.org/10.1007/s42484-021-00061-x}{Quantum Machine Intelligence \textbf{4(1)}, 3 (2022).}

\bibitem {ref30} Landman, Jonas and Mathur, Natansh and Li, Yun Yvonna and Strahm, Martin and Kazdaghli, Skander and Prakash, Anupam and Kerenidis, Iordanis."Quantum methods for neural networks and application to medical image classification." \href{https://doi.org/10.22331/q-2022-12-22-881}{Quantum \textbf{6}, 881(2022).}

\bibitem {ref31} Huang, Rui and Tan, Xiaoqing and Xu, Qingshan."Variational quantum tensor networks classifiers."  \href{https://doi.org/10.1016/j.neucom.2021.04.074}{Neurocomputing \textbf{452}, 89-98 (2021).}

\bibitem {ref32} Li, YaoChong and Zhou, Ri-Gui and Xu, RuQing and Luo, Jia and Hu, WenWen. "A quantum deep convolutional neural network for image recognition." \href{https://doi.org/10.1088/2058-9565/ab9f93}{Quantum Science and Technology\textbf{5(4)}, 044003 (2020).}

\bibitem {ref33} Song, Yanqi and Wu, Yusen and Wu, Shengyao and Li, Dandan and Wen, Qiaoyan and Qin, Sujuan and Gao, Fei. "A quantum federated learning framework for classical clients." \href{https://doi.org/10.1007/s11433-023-2337-2}{Science China Physics, Mechanics \& Astronomy\textbf{67(5)}, 250311 (2024).}

\bibitem {ref34} Stockwell, Robert G. "S-transform analysis of gravity wave activity from a small scale network of airglow imagers." \href{http://adsabs. harvard. edu/cgi-bin/nph-bib_query}{PhD thesis, The University of Western Ontario (1999).}

\bibitem {ref35} Crooks, Gavin E. "Gradients of parameterized quantum gates using the parameter-shift rule and gate decomposition." 
\href{https://doi.org/10.48550/arXiv.1905.13311}{arxiv preprint arxiv:1905.13311 (2019).}

\bibitem {ref36} Garcia, Carlos Iturrino and Grasso, Francesco and Luchetta, Antonio and Piccirilli, Maria Cristina and Paolucci, Libero and Talluri, Giacomo."A comparison of power quality disturbance detection and classification methods using CNN, LSTM and CNN-LSTM." 
\href{ https://doi.org/10.3390/app10196755}{Applied sciences\textbf{10(19)},6755(2020).}

\bibitem {ref37} Ahila, R and Sadasivam, V and Manimala, KJASC."An integrated PSO for parameter determination and feature selection of ELM and its application in classification of power system disturbances." 
\href{https://doi.org/10.1016/j.asoc.2015.03.036}{Applied Soft Computing\textbf{32},23-37(2015).}

\bibitem {ref38} Valtierra-Rodriguez, Martin and de Jesus Romero-Troncoso, Rene and Osornio-Rios, Roque Alfredo and Garcia-Perez, Arturo."Detection and Classification of Single and Combined Power Quality Disturbances Using Neural Networks." 
\href{https://doi.org/10.1109/TIE.2013.2272276}{IEEE Transactions on Industrial Electronics\textbf{61(5)},2473-2482(2014).}

\bibitem {ref39} Thirumala, Karthik and Pal, Sushmita and Jain, Trapti and Umarikar, Amod C."A classification method for multiple power quality disturbances using EWT based adaptive filtering and multiclass SVM." 
\href{https://doi.org/10.1016/j.neucom.2019.01.038}{Neurocomputing\textbf{334},265-274(2019).}

\bibitem {ref40} Gong, Renxi and Ruan, Taoyu."A new convolutional network structure for power quality disturbance identification and classification in micro-grids." 
\href{https://doi.org/10.1109/ACCESS.2020.2993202}{IEEE Access\textbf{ 8},88801-88814(2020).}

\bibitem {ref41} "IEEE Recommended Practice for Monitoring Electric Power Quality." 
\href{https://doi.org/10.1109/IEEESTD.2019.8796486}{IEEE Std 1159-2019 (Revision of IEEE Std 1159-2009),1-98(2019).}

\bibitem {ref42} Schuld, Maria and Bergholm, Ville and Gogolin, Christian and Izaac, Josh and Killoran, Nathan."Evaluating analytic gradients on quantum hardware."
\href{https://doi.org/10.1103/PhysRevA.99.032331}{Physical Review A\textbf{99(3)},032331(2019).}

\bibitem {ref43} Minh Khoa, Ngo and Van Dai, Le."Detection and classification of power quality disturbances in power system using modified-combination between the stockwell transform and decision tree methods."
\href{ https://doi.org/10.3390/en13143623}{Energies\textbf{13(14)},3623(2020).}

\bibitem {ref44} Cai, Kewei and Cao, Wenping and Aarniovuori, Lassi and Pang, Hongshuai and Lin, Yuanshan and Li, Guofeng."Classification of Power Quality Disturbances Using Wigner-Ville Distribution and Deep Convolutional Neural Networks."
\href{https://doi.org/10.1109/ACCESS.2019.2937193}{IEEE Access\textbf{7},119099-119109(2019).}

\bibitem {ref45} Montanaro, Ashley."Quantum algorithms: an overview."
\href{https://doi.org/10.1038/npjqi.2015.23}{npj Quantum Information\textbf{2(1)},1-8(2016).}

\bibitem {ref46} Childs, Andrew M."Lecture notes on quantum algorithms."
\href{https://doi.org/10.1109/ACCESS.2019.2937193}{Lecture notes at University of Maryland,114(2017).}

\bibitem {ref47} Stein, Samuel A and Baheri, Betis and Chen, Daniel and Mao, Ying and Guan, Qiang and Li, Ang and Xu, Shuai and Ding, Caiwen."Quclassi: A hybrid deep neural network architecture based on quantum state fidelity."
\href{https://proceedings.mlsys.org/paper_files/paper/2022/file/928f1160e52192e3e0017fb63ab65391-Paper.pdf}{Proceedings of Machine Learning and Systems\textbf{4},251-264(2022).}

\bibitem {ref48} Kyriienko, Oleksandr and Paine, Annie E and Elfving, Vincent E."Solving nonlinear differential equations with differentiable quantum circuits."
\href{https://doi.org/10.1103/PhysRevA.103.052416}{Physical Review A\textbf{103(5)},052416(2021).}

\bibitem {ref49} Nielsen, Michael A and Chuang, Isaac L."Quantum computation and quantum information."{Cambridge university press, (2010).}

\end{thebibliography}

\end{document}